\documentclass{nature}

\usepackage{amsmath, mathtools}
\usepackage{url}
\usepackage{xcolor} 
\usepackage{lineno}
\usepackage[normalem]{ulem}

\title{Vector-based Pedestrian Navigation in Cities}

\author{Christian Bongiorno$^{1,2}$, Yulun Zhou$^{1,3}$, Marta Kryven$^4$, David Theurel$^5$, Alessandro Rizzo$^6$, Paolo Santi$^{1,7}$, Joshua Tenenbaum$^4$ \& Carlo Ratti$^1$}

\begin{document}

\maketitle

\begin{affiliations}
 \item Senseable City Lab, Massachusetts Institute of Technology, 77 Mass Av, 02139, Cambridge, MA, US.
 \item Universit\'e Paris-Saclay, CentraleSup\'elec, Math\'ematiques et Informatique pour la Complexit\'e et les Syst\`emes, 91190, Gif-sur-Yvette, France.
 \item Department of Urban Planning and Design, Faculty of Architecture, The University of Hong Kong, Pokfulam, Hong Kong, China
 \item Department of Brain and Cognitive Sciences, Massachusetts Institute of Technology, 77 Mass Av, 02139, Cambridge, MA, US
 \item Department of Physics, Massachusetts Institute of Technology, 77 Mass Av, 02139, Cambridge, MA, US
 \item Dipartimento di Elettronica e Telecomunicazioni, Politecnico di Torino, 10128 Torino TO, Italy \\ Office of Innovation, New York University Tandon School of Engineering, Six MetroTech Center, Brooklyn NY 11201, US.
 \item Istituto di Informatica e Telematica del CNR, Via G. Moruzzi 1, 56124, Pisa, Italy.
\end{affiliations}

\begin{abstract}
How do pedestrians choose their paths within city street networks? Researchers have tried to shed light on this matter through strictly controlled experiments, but an ultimate answer based on real-world mobility data is still lacking. Here, we analyze salient features of human path planning through a statistical analysis of a massive dataset of GPS traces, which reveals that (1) people increasingly deviate from the shortest path when the distance between origin and destination increases, and (2) chosen paths are statistically different when origin and destination are swapped. We posit that direction to goal is a main driver of path planning and develop a vector-based navigation model that is a statistically better predictor of human paths than a model based on minimizing distance with stochastic effects. Our findings generalize across two major US cities with different street networks, hinting to the fact that vector-based navigation might be a universal property of human path planning.
\end{abstract}


\section{Introduction}
Although path planning is one of the hardest problems to solve computationally,~\cite{newell1958elements} humans plan remarkably efficient paths when navigating cities, and do so at various scales. While human path planning can be near-optimal, it also exhibits systematic divergences from the shortest available path,~\cite{zhu2015people}~\cite{lima2016understanding,javadi2017hippocampal} which are still not well understood. We hypothesise that a contribution to
such divergences arise from a common mental computational mechanism, which is shared between humans, can be modelled in precise quantitative terms, and generalises across urban environments.
Describing this mechanism by a formal computational account can help explain mental computations in the brain that support human mobility, inform the design of real-time planning tools that can better couple human and machine intelligence, and improve urban planning.

In the laboratory setting humans often rely on the \textit{approximate rationality principle}, meaning that humans use approximate planning heuristics to maximise their goals, while limiting subjective costs.~\cite{griffiths2015rational,huys2015interplay,gershman2015computational,baker2017rational,liu2017ten} In the real world these costs may be a combination of mental and physical effort~\cite{gershman2020origin}; for example, the mental cost of planning a route comes with the physical cost of travel.  
Multiple studies have investigated aggregate mobility flows~\cite{hillier2005network,brockmann2006scaling,gonzalez2008understanding,simini2012universal,alessandretti2018evidence,hamedmoghadam2019revealing,kraemer2020mapping,verbavatz2020growth,alessandretti2020scales,er2020universal,gallotti2016stochastic} and cognitive abilities that support navigation.~\cite{gillner1998navigation,foo2005humans,norman2005perception,sun2010perception,weisberg2016some,vuong2019no,becu2020age,van2020large} 
Quantitative models of how humans may plan their routes in real cities have also been proposed, although limited to a small neighborhood, in which the planning problem could be optimally solved with exhaustive route enumeration by Breadth First Search.\cite{javadi2017hippocampal}
However, generalizable computational models, that can generate precise quantitative predictions in large-scale city environments, are still lacking.
Thus, it is still unclear which subjective costs and planning heuristics may explain human routes in real urban environments. 


We investigate this question by analyzing a large dataset of GPS traces of 552,478 pseudo-anonymised human paths undertaken by 14,380 pedestrians in two major US cities -- Boston and San Francisco (see Methods). The original dataset was retrieved from a company used to run one of the largest mobile apps for mobility tracking. With full consent from users, the app records high-resolution trajectories of human movements in complex urban environments in their daily life. Activity types have been labeled by the company using proprietary machine learning tools, and only activities labeled as ``walking path" were analyzed in this work. Compared to previous studies of aggregate human mobility, which mostly rely on sparse location sampling -- such as the serial numbers of US banknotes,~\cite{brockmann2006scaling} surveys~,\cite{marshall2018mathematical,verbavatz2020growth}
commuting trips,~\cite{er2020universal} mobile communication records,~\cite{gonzalez2008understanding} social-media check-ins,~\cite{yan2017universal,yan2019destination} or location history ~\cite{alessandretti2018evidence,kraemer2020mapping} -- 
we study high resolution routes of individual pedestrians, reconstructed from their GPS traces. Thanks to pseudo-anonymisation, which assigns a unique anonymous ID to each individual in the data set, we could also associate multiple trajectories to the same individual, allowing  the study of individual-level properties of pedestrian routes -- see Supplementary, Section 5.
Importantly, the majority of human trajectories in our dataset were substantially different from routes suggested by Google Maps (see  Supplementary Tables 1, 2, 3 and Supplementary Figure 3), indicating a minimal bias introduced by machine-generated trajectories on our data. 
This property, along with recordings of multiple trips by the same individual, and between the same locations, enabled us to fit and evaluate alternative quantitative models of  pedestrian navigation. In particular, we restricted our attention to simple geometric, computational models that can be used to at least partially explain human navigation ability.

\begin{figure}
\centering
\includegraphics[width=1.3\textwidth]{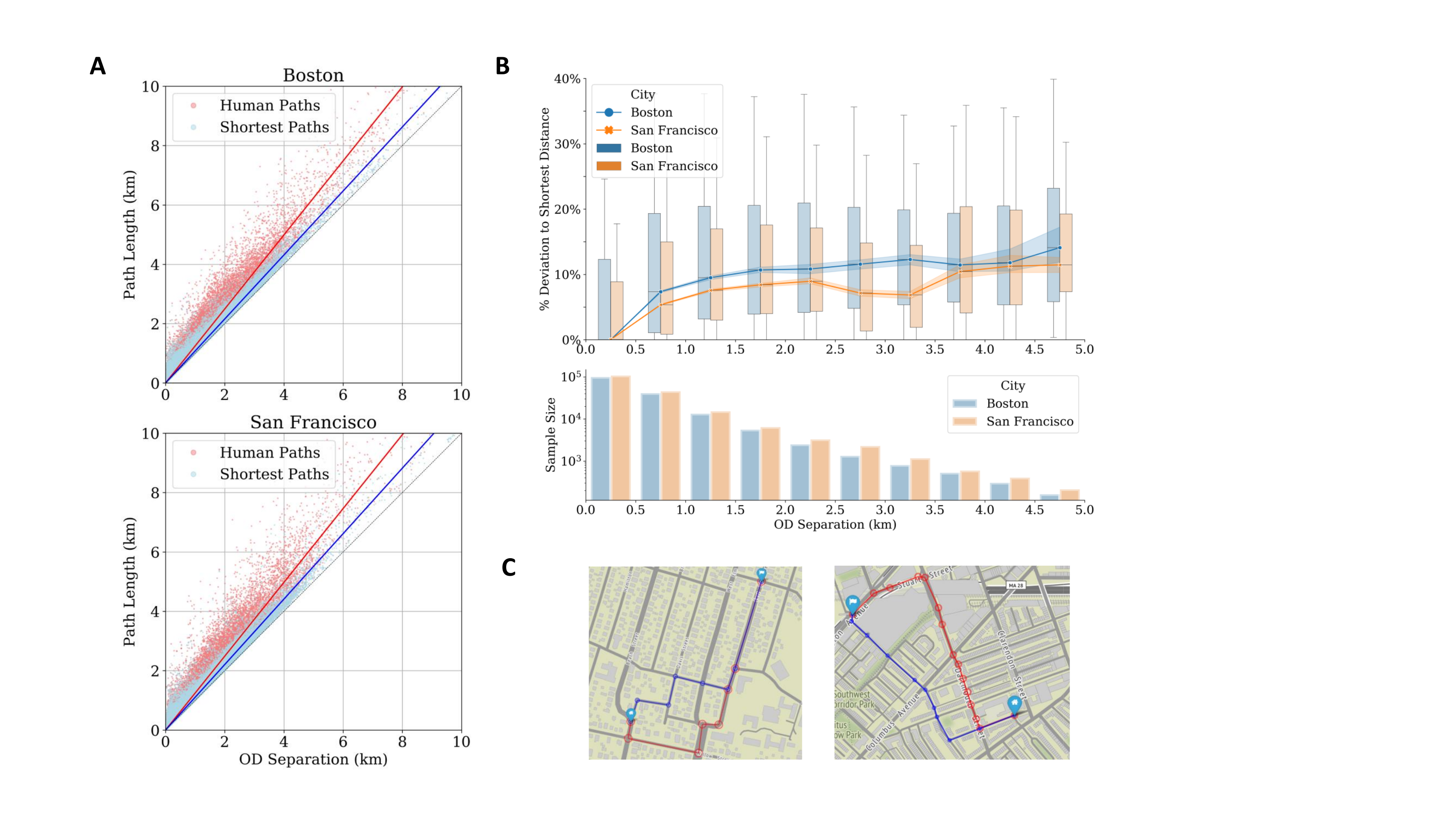}
\caption{\textbf{Differences between human paths and shortest paths.} A. Aggregated comparisons between the path lengths of human and shortest paths ($y$-axis) in Boston and San Francisco, as a function of the Euclidean distance between origin and destination  ($x$-axis); B. Relative differences between human and shortest path lengths ($y$-axis) as a function of the Euclidean distance between origin and destination  ($x$-axis); C. Two examples of the difference between human paths (in red) and their corresponding shortest paths (in blue).}
\label{fig:humanpath_are_different}
\end{figure}

\section{Results}
\subsection{Evaluating Paths Based on Distance}

As a first approximation, consider a simple way to formalise the cost of a path as equal to its distance: 
\begin{equation}
    C_{dist}[\mathcal{P}]=\sum_{S_i\in \mathcal{P}}l_{i}~.\nonumber
\end{equation}
Here, $\mathcal{P}$ denotes a path connecting the origin to the destination, composed of a list of street segments $S_1,S_2,\dots$, and $l_{i}$ denotes the walking distance along segment $S_i$. 
While this simple model captures the motivation to reduce distance, which is central to human planning~\cite{coutrot2019virtual,manley2015shortest}, 
several studies show that humans often deviate from the shortest paths \cite{hillier2005network,manley2015shortest,weisberg2016some,zhu2015people,Malleson2018}.

Indeed, Figure \ref{fig:humanpath_are_different}-A shows that human paths recorded in our dataset were consistently longer than the shortest distance path, computed using the standard Dijkstra algorithm \cite{Dijkstra1959}. 
The tendency to deviate from the shortest path increases with distance between origin and destination (see Figure \ref{fig:humanpath_are_different}-B), which could be due to the increasing complexity of evaluating relatively longer paths, in line with the approximate rationality principle. However, it is also interesting to observe that most of the relative deviation from shortest path is achieved by paths of length around 1~km, while only a modest further increasing deviation is observed for longer paths.

\subsection{Stochastic Distance Minimization}

The increasing deviation from the shortest path observed in the data could arise from uncertainty about the lengths of street segments, which leads to an accumulation of errors over time.  Formally, we can describe this process by an error term in the evaluation of street segment lengths:
\[
l_i\rightarrow e^{\mathcal{N}(\log(l_i),\sigma)}=c(l_i)~,
\]
and 
\begin{equation}\label{eq:distanceCost}
C_{dist}[\mathcal{P}]=\sum_{S_i\in \mathcal{P}}c(l_{i})~,
\end{equation}
where the new cost function $c(l_i)$ is obtained by applying a log-normally distributed random noise to the original street segment length $l_i$. The use of log-normal distribution to model uncertainty in street length estimation is motivated by the widely accepted Weber-Fechner law of just noticeable difference~\cite{Fechner1860}, which states that humans perceive measurable quantities on a logarithmic scale. We will refer to equation (\ref{eq:distanceCost}) as the {\em stochastic distance minimization} model. 
The number of street segments in the cost function (\ref{eq:distanceCost}) tends to increase with the Euclidean distance separating origin and destination. Hence, the deviation from the shortest distance tends to accumulate with increased separation between origin and destination, as we observed.

Importantly, in Methods we show that this model predicts path choices to be symmetrical -- the available paths are ranked in the same order of preference regardless of travel direction, which may not be the case in human paths. Navigation asymmetry has been demonstrated in the laboratory~\cite{newcombe1999misestimations}\cite{bailenson1998road,bailenson2000initial,christenfeld1995choices}, and empirically noted -- although never statistically tested -- in pedestrian flows~\cite{Malleson2018} and driving routes \cite{manley2015shortest}. A well-established theory of navigation by line of sight predicts asymmetric routes by suggesting that people travel along straight lines of sight in the desired direction, and when their view is obstructed, establish a new line of sight~\cite{hillier2005network}. 
Empirical findings in neuroscience~\cite{howard2014hippocampus,marchette2014anchoring} and psychology~\cite{newcombe1999misestimations,sun2010perception} further suggest that neural and mental representations of space and direction could lead to asymmetry in two ways. 
Landmark-based navigation, extensively documented in humans~\cite{gillner1998navigation,foo2005humans,sun2010perception,van2020large} and animals,~\cite{collett2004animal} may lead to asymmetric paths that depend on the distribution of views in the environment. Further, many animals, such as rodents,~\cite{hafting2005microstructure} bats,~\cite{de2017spatial,toledo2020cognitive} and cats~\cite{poucet1983route} rely on 
direction~\cite{hafting2005microstructure,howard2014hippocampus,marchette2014anchoring,epstein2017cognitive,poulter2020vector} for vector navigation\cite{poulter2020vector,tolman1948cognitive}. 
Human subjects in the laboratory studies likewise rely on direction for navigation, taking a route to the first destination that begins in the direction of subsequent destinations\cite{fu2015single}, and rely on neural representations of both Euclidian distance, and direction to destination\cite{howard2014hippocampus}. Several small-scale studies found that human routes exhibit systematic biases - such as bias toward paths that begin with a longer initial segment~\cite{bailenson2000initial,bailenson1998road,christenfeld1995choices}, or a southern bias~\cite{brunye2012planning}, which may contribute to asymmetries. However, such biases have not yet been computationally modeled, nor quantitatively evaluated, in large-scare urban environments.

\begin{figure}
    \centering
    \includegraphics[width=\textwidth]{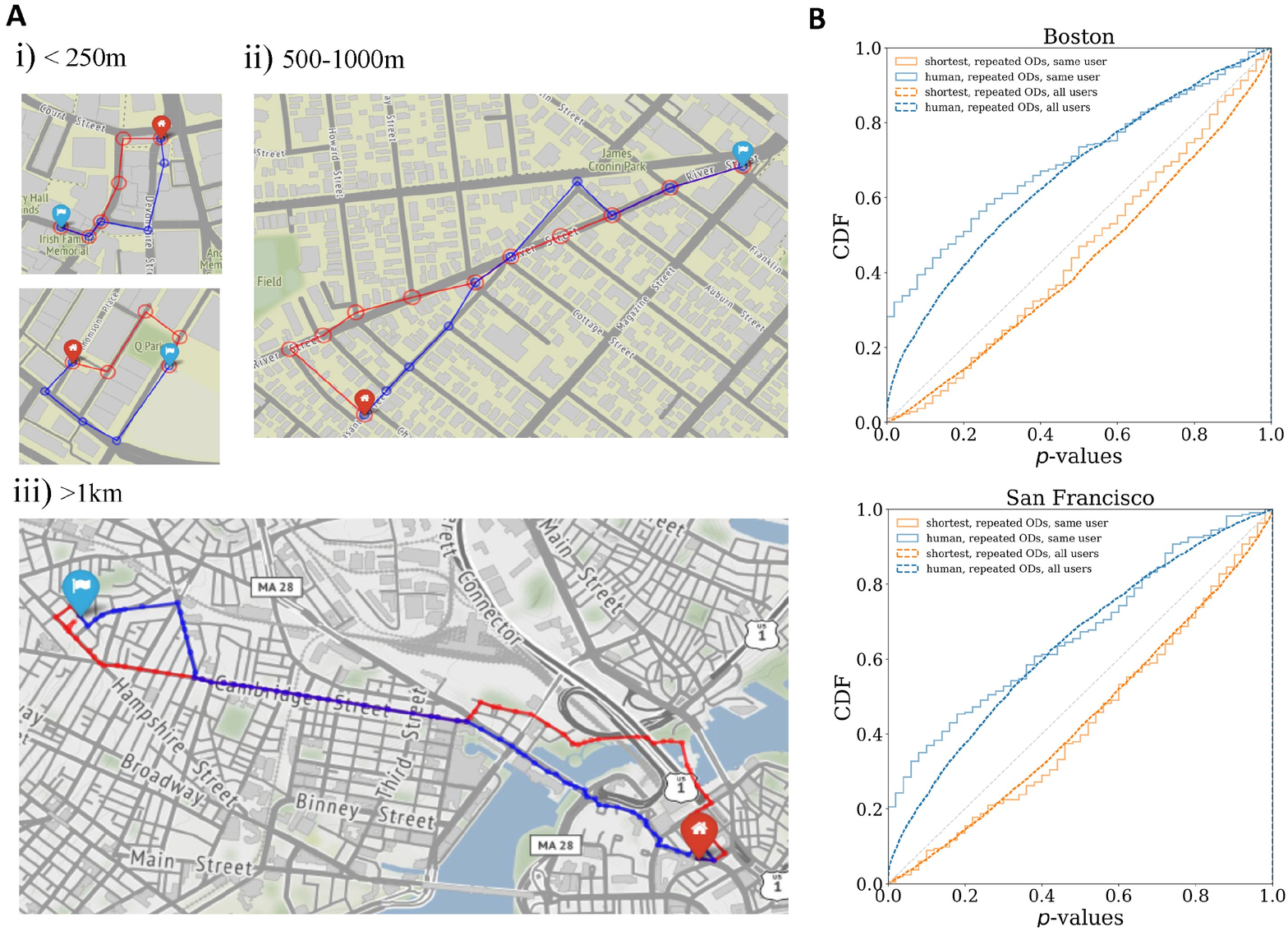}
    \caption{{\bf Asymmetric human paths.} A. Examples of asymmetric human paths in street networks. Red paths start at red markers, blue paths start at blue markers. B. Cumulative Distribution Function (CDF) of the one-tail p-values for the asymmetry test in Boston and San Francisco (method section). p-values being consistently above the diagonal dashed line indicates a statistically significant deviation from symmetric paths. Solid lines refer to individual-level tests; dotted lines refer to aggregate-level tests. Shortest distance paths used in the tests are obtained by simulating the stochastic distance minimization model.}
    \label{fig:asym}
\end{figure}

If present in human paths, asymmetric routes would falsify the stochastic distance minimization model as the only main contributor to pedestrian path formation. We tested for asymmetry as follows. First, we tested for asymmetric path choices within the same individual repeatedly visiting any given two locations in either order. If present, such individual-level asymmetries could arise from a cognitive cost heuristic used to evaluate trajectories, which depends on direction, or from stochasticity in distance evaluation, followed by memorising the planned routes. Second, to disambiguate these two scenarios, we tested for asymmetries in repeated paths aggregated over different pedestrians. Such an asymmetry would indicate a persistent quality of human path planning, which cannot be explained by stochastic route learning, and thus may be attributed to a common direction-dependent cost heuristic.
We have performed extensive statistical asymmetry tests at both individual and aggregate level (see Methods for details). The results of the test, reported in Figure~\ref{fig:asym}, showed a statistically significant asymmetry in individual walking trajectories, as well as in aggregate trajectories, suggesting that asymmetries are a persistent quality of pedestrian navigation in urban street networks (see Methods for details).

\subsection{Vector-based Navigation Model}

Other geometric models, such as the Initial Straightest Segment (ISS) strategy proposed by Bailenson et al. in \cite{bailenson2000initial}, could be used to explain the observed asymmetry in human paths.  If humans manifested a preference for the straightest first segment, we would expect to observe relatively longer first road segments in the human path than in the shortest path. However, human paths have consistently shorter initial segments than shortest paths, hinting to the fact that ISS strategy is likely not a cause of the observed asymmetry (see Supplementary Section 5.2). 

\begin{figure}
\centering

\includegraphics[width=\linewidth]{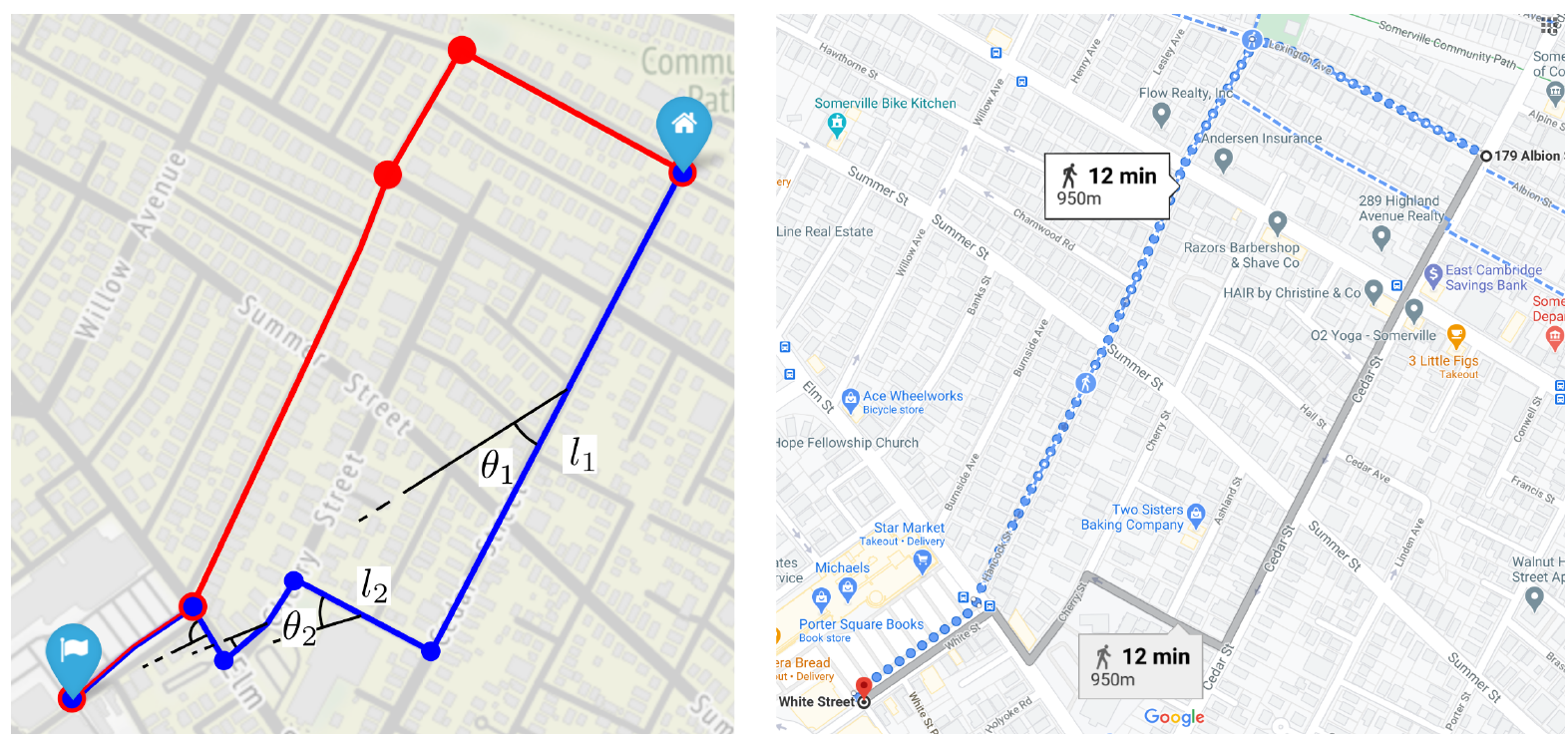}
\caption{{\bf Model design.} Example illustrating the calculation of the vector-based cost approximated by the Equation ~\eqref{eq:directionCost}. Left: the human path is shown in red; the shortest path is shown in blue. The quantities $l_i,\theta_i$ appearing in Equations~(\ref{eq:distanceCost},~\ref{eq:directionCost}) are shown for the first street segment of the blue path. Right: path alternatives produced by Google maps. (Note that Google reports both path lengths as 950m.)
}
\label{fig:NavModel}
\end{figure}

Then, inspired by the evidence of asymmetry in human paths, and the prevalence of vector-navigation in animal models, we hypothesised that humans use direction when planning their paths. 
We formalise the {\em vector-based navigation model} as a cost that depends on the angular deviation of the street segment from the destination: 
\begin{equation}\label{eq:directionCost}
C_{dir}[\mathcal{P}]=\sum_{S_i\in \mathcal{P}}c(\theta_{i},l_i)~,
\end{equation}
where
\[
c(\theta_i,l_i)=e^{\mathcal{N}(\log(|\theta_i|l_i),\sigma)}~,
\]
and $\theta_i\in[-\pi,\pi]$ represents the angle between the tangent to the path at street segment $S_i$ and the straight line to the destination (see Figure \ref{fig:NavModel}). In the example reported in Figure \ref{fig:NavModel}-a,  the stochastic distance minimization model using Equation~\eqref{eq:distanceCost} slightly prefers the blue path (for which $\sum l_i=935$m) over the red path ($\sum l_i=938$m). The vector-based navigation model prefers the leanest red path ($\sum|\theta_i|l_i=400$m$\cdot$rad) over the blue path ($\sum|\theta_i|l_i=516$m$\cdot$rad).
Since the angular deviations of street segments in this model depend on direction, the cost estimate of a path may change if its direction is reversed, implying that vector-based navigation could explain asymmetries observed in human trajectories (see Methods for a detailed proof.)

\subsection{Model comparison}
\begin{figure}
    \centering
    \includegraphics[width=\textwidth]{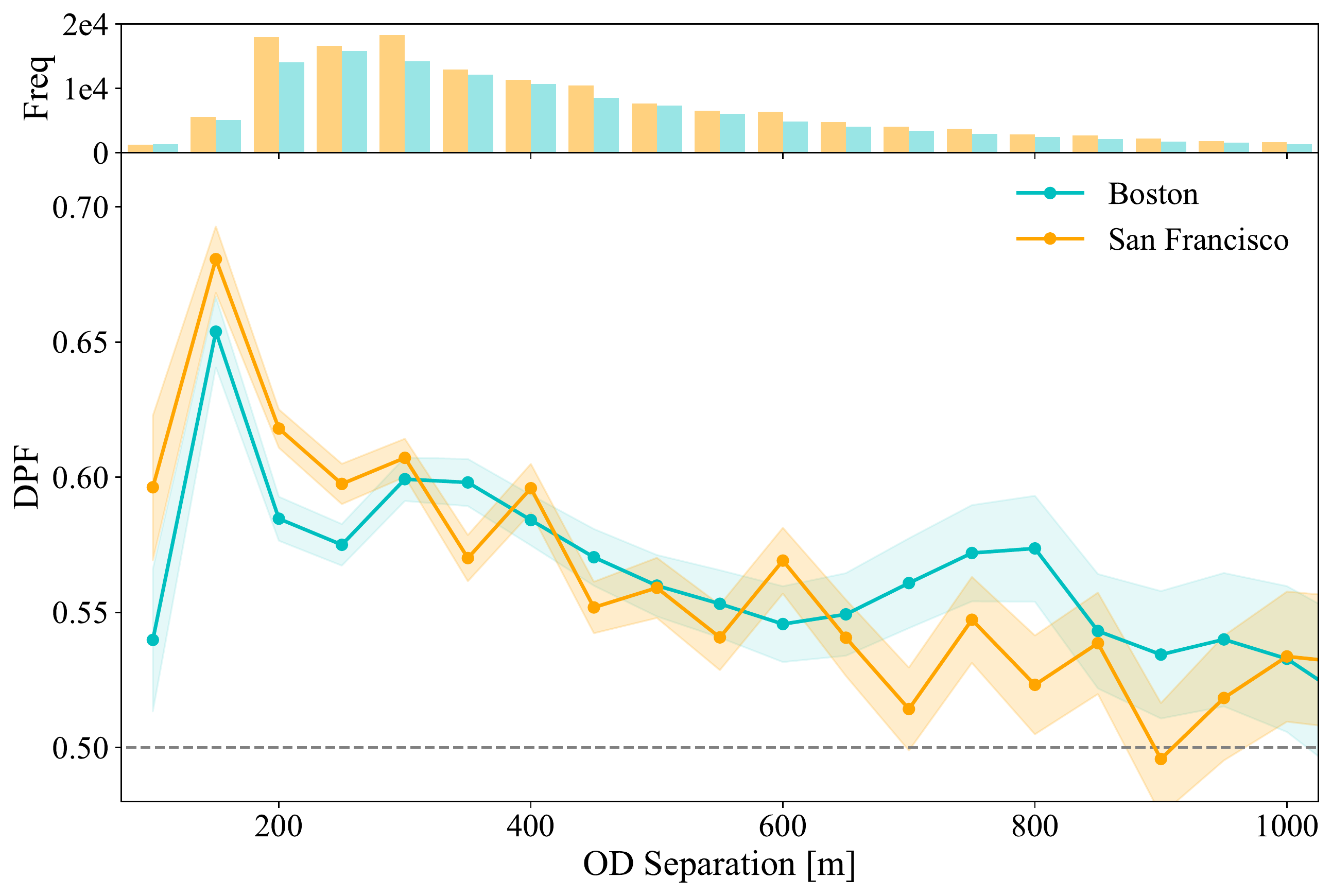}

    \caption{{\bf Directional Prevalence Fraction (DPF) for different OD separations.} Top: marginal frequency distributions of paths in Boston and San Francisco. Bottom: Similar DPF patterns in Boston and San Francisco. Error bars are obtained with the 95\% Wilson-score interval~\cite{wilson1927probable}.
    \label{fig:micro}}
\end{figure}

Having formally defined the two models, we compare their explanatory power relative to each other.
We aggregated all paths by OD distance separation, in steps of $50\mbox{m}$, and estimated the most likely parameters for the stochastic distance minimization and for the vector-based model in each bin (see Methods). 
We measure the fraction of paths in each bin for which the vector-based model has a higher point-wise likelihood than the stochastic distance minimization model -- we call this the {\em directional prevalence fraction} (DPF). A value higher than 50\% would imply a higher probability that a class of paths is explained through the vector-based model than the stochastic distance minimization model. The results reported in Figure~\ref{fig:micro} show that for each bin of OD distance 
separation with enough samples to reach statistical significance (see Supplementary Section 3)  the DPF is consistently above 50\%, suggesting the higher explanatory power of the vector-based model. The decreasing trend of the DPF function as OD distance 
separation increases can be consistently observed  both in Boston and in San Francisco. Moreover, the value of the DPF function is very similar across the entire range of 
OD distance 
separation, ranging from a peak of 68\% at 150~m separation -- which corresponds to 35\% better predictive power of vector-based vs. stochastic shortest distance model-- to 53\% at 1~Km separation. For the sake of completeness, the $\sigma$ parameters on both models reach overall different values, which are 0.44 and 1.06 for the stochastic vector-based and shortest path model in Boston, and 0.42 and 1.1 in San Francisco. More details on the calibration process and the detailed shape of the $\sigma$ function are reported in the Supplementary Figures 4 and 5.

The decreasing trend of the DPF function suggests a metacognitive mental computational mechanism, which trades off mental and physical costs during navigation. As the length of the planned path increases, humans likely shift from the relatively easier direction optimization strategy toward optimizing distance, in order to save physical effort and travel time. This hypothesis is also supported by the observed deviation of human trajectories from shorter paths, which tends to level-off for paths longer than 1~km -- recall Figure \ref{fig:humanpath_are_different}. Observing this trend in two cities with drastically different street network topology suggests that this mental process generalises across city environments.

\section{Discussion}

The data-driven approach used in this paper illustrates how large-scale observations of behaviour in natural settings can be used to derive quantitative models of complex cognitive and physical tasks, and complements studies of cognition in the laboratory by providing unique insight into how the different cognitive faculties work together in real life. Despite all the challenges of an uncontrolled real city environment, as well as lack of information about users and their activities that limits the range of applicability of our findings, we have discovered behavioural trends which generalise across cities, attesting to the  potential of big-data in the fields of psychology and cognitive science. 

Our results suggest that vector navigation may be a common property of human route planning in cities, and establish a direct connection between the study of human mobility, human cognitive psychology, and neuroscience.
The empirical support of the vector-navigation model by real human behaviour provides evidence in favour of theoretical claims that human minds and brains represent both Euclidean maps and graphs of street networks, as suggested in prior work\cite{peer2020structuring}. 
The quantitative cost models described in this work suggest a possible computational mechanism by which route planning may be implemented in the brain. However, more work is needed to overcome the limitations of our study -- lack of information about users and their activity -- and understand, for instance, the effect of individual differences, the extent to which people pre-plan their routes, use hierarchical map representations, develop routing habits, and diversify path alternatives as a result of map learning.

The requirement to implement precise falsifiable models has limited our study to factors that could be measured objectively and exactly -- such as direction and distance -- as well as a number of simplifying assumptions.
While our models assume that pedestrians have accurate map knowledge, future work should consider how human path planning may depend on mental representations of the street networks, which may include larger streets and omit secondary ones\cite{stern1988levels}. Future work should also consider other factors that can influence navigation decisions, to the extent that these factors can be quantitatively tracked -- such as day of the week, sunlight, weather, trees, attractions, presence of crowds, fatigue, time of the day, neighborhood safety, and elevation gradient -- as well as individual differences in the responsiveness to these factors.

Our results extend our understanding of human route planning, and could have significant implications for those areas where route planning is a basic mechanism, such  as transportation, cognitive maps, and real-time planning algorithms.
We expect our methodology to become of even greater use with increasingly accurate 5G and 6G tracking data. The latter should allow monitoring which side of the street someone is walking, the time spent at the intersections, forward facing direction, and speed, supporting a more detailed investigation of the internal planning mechanisms.  
Future models should account for walking speed, tendency to change direction at a red light, as well as tendency to consistently orient toward specific visual landmarks. 
We hope that our findings will stimulate new research exploring these connections.

\section*{Methods}
\subsection{Dataset description and preparation}
The full dataset comprises 579,231 pseudo-anonymised human paths produced by 14,380 pedestrians, 5,590 in Boston and 8,790 in San Francisco, recorded by an always-on pedestrian tracking smartphone application over a time-period of one year. As customary in research based on data sets acquired through mobile applications, we do not have information about the demographics of the population of app users, which is then not necessarily representing an unbiased sample of the population in Boston and San Francisco. The two cities differ in street network geometry (see Supplementary Figure 1). 
 San Francisco street network is designed as a grid, while Boston street network is highly irregular for historical reasons. The application was continuously recording individuals' movements throughout the day. Thus, the dataset consists of raw high-quality GPS traces, which include a range of pedestrian activities. The analysis reported in the paper focuses on activities that has been labeled as ``walking activity" by the app using proprietary machine learning algorithms. Also, due to privacy limitations, we have no information about user profiles, such whether they are resident of the city, level of familiarity with the environment, etc. The average recording gap is 15s, and the positioning accuracy is within 10m. The GPS traces were segmented to individual paths based on tracking continuity - a path is considered to end at a destination if the walking activity was paused for longer than 5 minutes. To preserve privacy, the origin and destination of each trip have been randomly relocated within a 100 meter radius centered at the original location. To remove any possible bias resulting from this randomisation procedure, we trimmed the beginning and the end of each path within this range.
All paths were map-matched to the open street map network available at {\tt www.openstreetmap.org} using a Hidden-Markov-Chain algorithm \cite{HMMmapmatching}. Furthermore, we screened all map-matched paths using the Douglas-Peucker (DP) algorithm~\cite{DPAlgorithm} to exclude any detours that could be caused by the GPS jitters, that were not fully removed by the trimming.

We analyzed a subset of walking paths selected from the full dataset, that met the following criteria: (1) no straight line paths, and (2) a paths was not more than 80\% longer than the shortest possible path connecting the origin and destination. The 80\% cutoff was taken as the 95\% quantile. The first criterion removed straightforward (from a navigational perspective) paths. The second criterion was implemented to exclude multi-purpose trips, sight-seeing, and exercise. We also excluded paths having their shortest network distance smaller than 200 meters. 
After this pre-processing 165,645 trajectories by 4,879 pedestrians from Boston, and 189,075 trajectories by 7,372 pedestrians from San Francisco remained in the analysis, comprised of about 60\% of paths from the original dataset.

The paths in Boston has a mean length of $856.0m$ ($SD=843.6m$), while the paths in San Francisco had a mean length of $868.1m$ ($SD =912.1m$). Detailed aggregate statistics of the human paths included in the study are reported in the Supplementary  Table 1.  

\subsection{Street Networks}
We retrieved the street network of the city of Boston and San Francisco and their surrounding areas from the Open Street Map (https://www.openstreetmap.org) (see Supplementary Figure 1). All walkable street segments are included to form the walkable street network that is used consistently for all the following calculation and analysis, including but not limited to the map-matching of human path, calculation of shortest path, and random walk paths. 

We simplified the retrieved street networks to speed up the calculation by cleaning up redundant nodes and edges around intersections. Specifically, we grouped adjacent intersections into one by applying a hierarchical clustering using the complete linkage and network distance among nodes. We selected 30m as the threshold of the diameter of the clusters through repeated experiments. This simplification eliminated unnecessary details around intersections while preserving network topology.  The street network of Boston and San Francisco used in our data analysis is illustrated in Supplementary Figure 1.

\subsection{Walking Paths}

The GPS trajectories of human paths were map-matched to the walking street network from the Open Street Map (Supplementary Section 1.1) using a Hidden Markov Chain algorithm (Newson and Krumm, 2009). The algorithm was implemented using the map matching API of Graphhoper (https://graphhopper.com/api/1/docs/map-matching/). A small number of paths that were not successfully matched were eliminated from the data-set. The map-matched human paths were projected onto the simplified network utilizing the link table between nodes in the original network and the simplified network (Supplementary Section 1.1). The summary statistics of the shortest paths in both Boston and San Francisco are shown in the Supplementary Table 1. 

For each human path, the corresponding shortest distance path was calculated based on the street network from the Open Street Map (Supplementary Section 1.1) using the classic Dijkstra's algorithm implemented in the python package of igraph-python  (http://igraph.org/python/). The map-matched shortest paths were projected onto the simplified network utilizing the link table between nodes in the original network and the simplified network. The summary statistics of the shortest paths in both Boston and San Francisco are shown in the Supplementary Table 1. The trip velocity distribution is reported in Supplementary Figure 2.

The Google paths were retrieved from the Google Map Routing API (https://cloud.google.com/maps-platform/routes/) and map-matched to the original street network using the identical HMM algorithm as applied to the human paths. Due to resource limitations, we randomly extracted a subset of paths for the comparison between human and google paths in Boston and San Francisco respectively. We retrieved Google paths for 9,254 origin-destination pairs and 1,492 origin-destination pairs in San Francisco. The comparison was restricted to the human paths for which we could obtain a corresponding Google path between the given origin and destination. If multiple human paths existed for one OD pair, all human paths were compared separately to the Google path, and average similarity performances were taken for the comparison. All retrieved Google trajectories were map-matched to the original street networks before being projected to the simplified network. 

To examine the probability of pedestrians following app-planned paths, we compared human paths to paths planned by the most widely used routing app - Google Map. Results show that Google paths are significantly different from human paths. First, the length distribution of Google path is significantly different from that of human paths in both cities (see Supplementary Figure 3). Metrics comparing geometries of Google paths with geometries of human paths also demonstrate a significant difference in Jaccard Similarity measured as exact overlaps, and in geometric similarity measured using Hausdorff distance (see Supplementary Tables 2 and 3).

\subsection{Detection of Asymmetries}

To test whether the sample probability distribution of paths chosen between any specific origin-destination (OD) pair depends on the direction  we need to aggregate trips across OD pairs. However, since the sample size obtained by considering the original OD pair is limited, we also consider intermediate points in a trajectory as possible OD pairs. 
To achieve a sufficiently large sample size we consider only OD pairs within an Euclidean separation distance between 200m and 250m, with at least 50 recorded paths, including at least 20 paths in each direction, that do not lie on a straight line path.  
We define a straight line path between an OD if there exists a path that simplifies to two points with the DP algorithm~\cite{DPAlgorithm}, with a cutoff of 30m. Note that this approach is conservative, since a symmetric model must necessarily be symmetric at all scales.

For individual-level asymmetry test this criterion is met by 235 OD pairs generated by 73 pedestrians in San Francisco, and by 316 OD pairs generated by 101 pedestrians in Boston. For the aggregate asymmetry analysis we also include any OD pairs that have a sufficient number of paths combined between different pedestrians, which gives 2,865 and 3,000 origin-destination pairs in San Francisco and Boston, respectively. Note that a single pedestrian path could be accounted multiple times by considering different sub-parts of the original trajectory.

Aggregate-level analysis is done as follows. For a specific OD pair, we consider the universe of all paths followed by humans in each direction, and represent them as an unordered set of street intersection points. For each intersection point, we record the occurrence of out-bound $OD^\rightarrow = \{n_1^\rightarrow, \cdots, n_m^\rightarrow \}$, and in-bound $OD^\leftarrow = \{n_1^\leftarrow, \cdots, n_m^\leftarrow \}$ paths. 
Individual-level analysis is done as follows. We group  all paths taken by an individual between a specific OD pair $OD=\{\mathcal{P}_1,\dots,\mathcal{P}_m\}$, to determine the universe of observed paths in each direction. We count the number of occurrences of these paths in the forward and reverse direction, defined as  $OD^\rightarrow=\{n^\rightarrow_1,\dots,n^\rightarrow_m\}$ and $OD^\leftarrow=\{n^\leftarrow_1,\dots,n^\leftarrow_m\}$, and establish an analogy with the process of extracting with replacement marbles from an urn containing $m$ different types of marbles -- one for each possible path. More specifically, the urn contains $n_i^\rightarrow+n_i^\leftarrow$ marbles of type $i$, for each $i$. Given this analogy, the observed group of paths in one direction, say, $OD^\rightarrow$, can be seen as an instance of a random extraction of $q^\rightarrow=\sum_i n_i^\rightarrow$ marbles from the urn. If paths were to be symmetric -- null hypothesis, the extraction would obey a multinomial distribution. Thus, we can apply standard statistical hypothesis test to the null hypothesis that the observations from sets $OD^\rightarrow$ and $OD^\leftarrow$ follows such distribution. 
Results of the test, reported in Figure~\ref{fig:asym}, show that the null hypothesis can be rejected with a 0.05 significance threshold for 32\% and 24\% of the origin-destination pairs for Boston and San Francisco respectively, and that the cumulative distribution of the resulting p-values is considerably skewed towards lower values than those expected under the null hypothesis. 
Thus, we find  a statistically significant asymmetry in individual human paths.

The proposed statistical test, which applies to both individual- and aggregate-level analysis, builds on the analogy with the process of independently extracting $m$ marbles from an urn. If the null hypothesis that paths are symmetric is true,  there exists a set of probabilities  $\{p^1, \cdots, p^m\}$ with $\sum_{i=1}^m p^i = 1$, associated to each path that are direction independent. In such a case, the best estimates for such probabilities can be obtained as 
\begin{equation}
    p^i = \frac{n_i^\leftarrow + n_i^\rightarrow}{q^\leftarrow + q^\rightarrow},
\end{equation}
with $q^\leftarrow = \sum_{i=1}^m n_i^\leftarrow$ and $q^\rightarrow = \sum_{i=1}^m n_i^\rightarrow$.

To assess the validity of the null hypothesis, we test whether the sample occurrences $OD^\rightarrow$ and $OD^\leftarrow$ are statistically compatible with the process of repeatedly and independently extracting $q^\rightarrow+ q^\leftarrow$ elements from a urn with $m$ marbles (with replacement). The resulting distribution is multinomial and defined by the probabilities $p_i$. The exact p-value for the test can not be obtained analytically; however, the Likelihood-Ratio test~\cite{wilks1938large} 
provides an asymptotic distribution in the case the null hypothesis holds. Specifically, the maximum likelihood estimate is
\begin{equation}
    \log \mathcal{L}_M = \sum_{i=1}^m n_i^\leftarrow \log \left(\frac{n_i^\leftarrow}{q^\leftarrow} \right) + \sum_{i=1}^m n_i^\rightarrow \log \left(\frac{n_i^\rightarrow}{q^\rightarrow} \right)
\end{equation}
and the alternative model is 
\begin{equation}
    \log \mathcal{L}_A = \sum_{i=1}^m n_i^\leftarrow \log(p_i) + \sum_{i=1}^m n_i^\rightarrow \log(p_i).
\end{equation}
Finally, according to Wilks’ theorem~\cite{wilks1938large}, the Likelihood-Ratio statistic should converge to a chi square with $m-1$ degrees of freedom
\begin{equation}
    2( \log \mathcal{L}_M - \log \mathcal{L}_A ) \xrightarrow[]{d}
\chi^2( m-1 ).
\end{equation} 

The results of the test are reported in the main text. Since the underlying distribution on which performing the statistical test is unknown, we proceed with a more conservative test based on a null sampled distribution. This implies that the real null expectation should be slightly below the first bisector, as depicted by the shortest distance p-value distribution in Figure \ref{fig:asym}-B. 

\subsection{Asymmetry Proofs}

To prove that the stochastic distance minimization model cannot explain asymmetry,  
consider the (random) cost $\mathcal{C}_{dist}(OD^\rightarrow)$ of a path between a certain origin-destination pair $OD$, and the cost $\mathcal{C}_{dist}(OD^\leftarrow)$ of the reverse path. Since both costs are obtained as sum of independent random variables, and the random variables considered in the summation are the same (except for their order), both $\mathcal{C}_{dist}(OD^\rightarrow)$ and $\mathcal{C}_{dist}(OD^\leftarrow)$ have the same probability distribution, which contradicts our empirical observations of asymmetric human paths.  

To prove that the vector-based navigation model can explain asymmetry, notice that, for a given path between O and D 
the cost of the path in the forward and reverse direction, $C_{dir
}(OD^{\rightarrow})$ and $C_{dir}(OD^\leftarrow)$, are obtained from the summation of {\em different} random variables, since the value $\theta_i^\rightarrow$ for street segment $S_i$ in the forward direction is different from that of $\theta_i^\leftarrow$ in the reverse direction -- see Figure \ref{fig:NavModel}. Thus, the probability distribution of $C_{dir
}(OD^\rightarrow)$ is different from that of $C_{dir}(OD^\leftarrow)$, which implies that the stochastic distance minimization model can support the hypothesis of asymmetric human navigation.

\subsection{Model Fitting and Comparison}

We compare the explanatory power of the two models by performing a set of $1000$ simulations in Boston and San Francisco to optimally tune the error parameter $\sigma$ for both the stochastic distance minimization and the vector-based model -- details about the exploration ranges for $\sigma$ can be found in the Supplementary Section 3.
Given a human path $OD^{{(h)}\rightarrow}$ from O to D, we run both models on that origin destination pair for $1000$ simulations, and recorded the number of times each of the two models selected exactly path $OD^{(h)\rightarrow}$.
To determine which of the two models performed statistically better in this task, we compared their respective likelihood values. The likelihood of a model is obtained from

\begin{equation}\nonumber
    \log \mathcal{L}^{(x)} = \sum_{i=1}^N \log P(OD^{(h)\rightarrow}|OD_i,C_{(x)},\sigma_x),
\end{equation}
that is, the sum over all $N$ paths of the logarithm of the probability $$P(OD^{(h)\rightarrow}|OD_i,C_{(x)},\sigma_x)$$ of selecting the human path $OD^{(h)\rightarrow}$ for origin-destination pair $OD_i$, computed using the cost function of the specific model $x=\{dist,dir\}$ with the optimal value $\sigma_x$ for model $x$ of the error parameter.

We aggregated paths into bins by OD separation, in steps of $50\mbox{m}$. For each bin we calibrated $\sigma$ and tested model performances with a leave-one-out cross-validation~\cite{zhang1993model}. So, we found the value of $\sigma$ that maximises the likelihood of $N_s-1$ paths associated with the bin $s$, and measured the likelihood of the out-of-sample path. We repeated this procedure by leaving out another path until we covered the whole set of paths in each bin. We also included a cutoff parameter $c$ to account for paths with zero sample-probability, which would cause divergence of the likelihood metric. Such parameter is set to $c=1/N_s=0.001$ which is the expected minimal detectable probability with $N_s=1000$ simulations. A detailed exploration of the dependency of $c$ is reported in the Supplementary Section 3, however we did not observe substantial change of our results with different values of $c$.

\section*{Data Availability Statement}
Due to privacy constrains policies and signed Data Usage Agreement, we are not allowed to share the full GPS tracks considered in this work. For this reason, we generated a small sample of 100 trajectories for Boston. We also make available the pre-processed pedestrian street networks for Boston and San Francisco. The sample data set and street network data can be accessed at Zenodo \cite{ZenodoDOI}. 

Source data for figures 1, 2, and 4 is also available with this manuscript. Figure 1c, 2A, and 3 (left) used basemap from Open Street Map (https://www.openstreetmap.org) under the Open Database License (https://www.openstreetmap.org/copyright). Figure 3 (right) uses Map data ©2021 Google under the fair-use guidelines (https://about.google/brand-resource-center/products-and-services/geo-guidelines/$\sharp$general-guidelines-copyright-fair-use).

\section*{Code Availability Statement}

The version of PedNav package used in this study and a guide to reproducing the results is available through GitHub under the GNU GPL-3.0 License ( https://github.com/cbongiorno/pednav). The specific version of the package used to generate the results in the current study can be found at Zenodo \cite{ZenodoDOI}.

A pseudo-code description of the algorithms used for human navigation based on stochastic distance minimization and vector navigation is reported in the Supplementary Section 4. 

\section{References}
\bibliographystyle{naturemag}
\bibliography{main.bib}

\begin{thebibliography}{10}
\expandafter\ifx\csname url\endcsname\relax
  \def\url#1{\texttt{#1}}\fi
\expandafter\ifx\csname urlprefix\endcsname\relax\def\urlprefix{URL }\fi
\providecommand{\bibinfo}[2]{#2}
\providecommand{\eprint}[2][]{\url{#2}}

\bibitem{newell1958elements}
\bibinfo{author}{Newell, A.}, \bibinfo{author}{Shaw, J.~C.} \&
  \bibinfo{author}{Simon, H.~A.}
\newblock \bibinfo{title}{Elements of a theory of human problem solving.}
\newblock \emph{\bibinfo{journal}{Psychological review}}
  \textbf{\bibinfo{volume}{65}}, \bibinfo{pages}{151} (\bibinfo{year}{1958}).

\bibitem{zhu2015people}
\bibinfo{author}{Zhu, S.} \& \bibinfo{author}{Levinson, D.}
\newblock \bibinfo{title}{Do people use the shortest path? an empirical test of
  wardrop’s first principle}.
\newblock \emph{\bibinfo{journal}{PloS one}} \textbf{\bibinfo{volume}{10}}
  (\bibinfo{year}{2015}).

\bibitem{lima2016understanding}
\bibinfo{author}{Lima, A.}, \bibinfo{author}{Stanojevic, R.},
  \bibinfo{author}{Papagiannaki, D.}, \bibinfo{author}{Rodriguez, P.} \&
  \bibinfo{author}{Gonz{\'a}lez, M.~C.}
\newblock \bibinfo{title}{Understanding individual routing behaviour}.
\newblock \emph{\bibinfo{journal}{Journal of The Royal Society Interface}}
  \textbf{\bibinfo{volume}{13}}, \bibinfo{pages}{20160021}
  (\bibinfo{year}{2016}).

\bibitem{javadi2017hippocampal}
\bibinfo{author}{Javadi, A.-H.} \emph{et~al.}
\newblock \bibinfo{title}{Hippocampal and prefrontal processing of network
  topology to simulate the future}.
\newblock \emph{\bibinfo{journal}{Nature communications}}
  \textbf{\bibinfo{volume}{8}}, \bibinfo{pages}{1--11} (\bibinfo{year}{2017}).

\bibitem{griffiths2015rational}
\bibinfo{author}{Griffiths, T.~L.}, \bibinfo{author}{Lieder, F.} \&
  \bibinfo{author}{Goodman, N.~D.}
\newblock \bibinfo{title}{Rational use of cognitive resources: Levels of
  analysis between the computational and the algorithmic}.
\newblock \emph{\bibinfo{journal}{Topics in cognitive science}}
  \textbf{\bibinfo{volume}{7}}, \bibinfo{pages}{217--229}
  (\bibinfo{year}{2015}).

\bibitem{huys2015interplay}
\bibinfo{author}{Huys, Q.~J.} \emph{et~al.}
\newblock \bibinfo{title}{Interplay of approximate planning strategies}.
\newblock \emph{\bibinfo{journal}{Proceedings of the National Academy of
  Sciences}} \textbf{\bibinfo{volume}{112}}, \bibinfo{pages}{3098--3103}
  (\bibinfo{year}{2015}).

\bibitem{gershman2015computational}
\bibinfo{author}{Gershman, S.~J.}, \bibinfo{author}{Horvitz, E.~J.} \&
  \bibinfo{author}{Tenenbaum, J.~B.}
\newblock \bibinfo{title}{Computational rationality: A converging paradigm for
  intelligence in brains, minds, and machines}.
\newblock \emph{\bibinfo{journal}{Science}} \textbf{\bibinfo{volume}{349}},
  \bibinfo{pages}{273--278} (\bibinfo{year}{2015}).

\bibitem{baker2017rational}
\bibinfo{author}{Baker, C.~L.}, \bibinfo{author}{Jara-Ettinger, J.},
  \bibinfo{author}{Saxe, R.} \& \bibinfo{author}{Tenenbaum, J.~B.}
\newblock \bibinfo{title}{Rational quantitative attribution of beliefs, desires
  and percepts in human mentalizing}.
\newblock \emph{\bibinfo{journal}{Nature Human Behaviour}}
  \textbf{\bibinfo{volume}{1}}, \bibinfo{pages}{1--10} (\bibinfo{year}{2017}).

\bibitem{liu2017ten}
\bibinfo{author}{Liu, S.}, \bibinfo{author}{Ullman, T.~D.},
  \bibinfo{author}{Tenenbaum, J.~B.} \& \bibinfo{author}{Spelke, E.~S.}
\newblock \bibinfo{title}{Ten-month-old infants infer the value of goals from
  the costs of actions}.
\newblock \emph{\bibinfo{journal}{Science}} \textbf{\bibinfo{volume}{358}},
  \bibinfo{pages}{1038--1041} (\bibinfo{year}{2017}).

\bibitem{gershman2020origin}
\bibinfo{author}{Gershman, S.~J.}
\newblock \bibinfo{title}{Origin of perseveration in the trade-off between
  reward and complexity}.
\newblock \emph{\bibinfo{journal}{bioRxiv}}  (\bibinfo{year}{2020}).

\bibitem{hillier2005network}
\bibinfo{author}{Hillier, B.} \& \bibinfo{author}{Iida, S.}
\newblock \bibinfo{title}{Network and psychological effects in urban movement}.
\newblock In \emph{\bibinfo{booktitle}{International Conference on Spatial
  Information Theory}}, \bibinfo{pages}{475--490}
  (\bibinfo{organization}{Springer}, \bibinfo{year}{2005}).

\bibitem{brockmann2006scaling}
\bibinfo{author}{Brockmann, D.}, \bibinfo{author}{Hufnagel, L.} \&
  \bibinfo{author}{Geisel, T.}
\newblock \bibinfo{title}{The scaling laws of human travel}.
\newblock \emph{\bibinfo{journal}{Nature}} \textbf{\bibinfo{volume}{439}},
  \bibinfo{pages}{462--465} (\bibinfo{year}{2006}).

\bibitem{gonzalez2008understanding}
\bibinfo{author}{Gonzalez, M.~C.}, \bibinfo{author}{Hidalgo, C.~A.} \&
  \bibinfo{author}{Barabasi, A.-L.}
\newblock \bibinfo{title}{Understanding individual human mobility patterns}.
\newblock \emph{\bibinfo{journal}{nature}} \textbf{\bibinfo{volume}{453}},
  \bibinfo{pages}{779--782} (\bibinfo{year}{2008}).

\bibitem{simini2012universal}
\bibinfo{author}{Simini, F.}, \bibinfo{author}{Gonz{\'a}lez, M.~C.},
  \bibinfo{author}{Maritan, A.} \& \bibinfo{author}{Barab{\'a}si, A.-L.}
\newblock \bibinfo{title}{A universal model for mobility and migration
  patterns}.
\newblock \emph{\bibinfo{journal}{Nature}} \textbf{\bibinfo{volume}{484}},
  \bibinfo{pages}{96--100} (\bibinfo{year}{2012}).

\bibitem{alessandretti2018evidence}
\bibinfo{author}{Alessandretti, L.}, \bibinfo{author}{Sapiezynski, P.},
  \bibinfo{author}{Sekara, V.}, \bibinfo{author}{Lehmann, S.} \&
  \bibinfo{author}{Baronchelli, A.}
\newblock \bibinfo{title}{Evidence for a conserved quantity in human mobility}.
\newblock \emph{\bibinfo{journal}{Nature Human Behaviour}}
  \textbf{\bibinfo{volume}{2}}, \bibinfo{pages}{485--491}
  (\bibinfo{year}{2018}).

\bibitem{hamedmoghadam2019revealing}
\bibinfo{author}{Hamedmoghadam, H.}, \bibinfo{author}{Ramezani, M.} \&
  \bibinfo{author}{Saberi, M.}
\newblock \bibinfo{title}{Revealing latent characteristics of mobility networks
  with coarse-graining}.
\newblock \emph{\bibinfo{journal}{Scientific reports}}
  \textbf{\bibinfo{volume}{9}}, \bibinfo{pages}{1--10} (\bibinfo{year}{2019}).

\bibitem{kraemer2020mapping}
\bibinfo{author}{Kraemer, M.~U.} \emph{et~al.}
\newblock \bibinfo{title}{Mapping global variation in human mobility}.
\newblock \emph{\bibinfo{journal}{Nature Human Behaviour}}
  \bibinfo{pages}{1--11} (\bibinfo{year}{2020}).

\bibitem{verbavatz2020growth}
\bibinfo{author}{Verbavatz, V.} \& \bibinfo{author}{Barthelemy, M.}
\newblock \bibinfo{title}{The growth equation of cities}.
\newblock \emph{\bibinfo{journal}{Nature}} \textbf{\bibinfo{volume}{587}},
  \bibinfo{pages}{397--401} (\bibinfo{year}{2020}).

\bibitem{alessandretti2020scales}
\bibinfo{author}{Alessandretti, L.}, \bibinfo{author}{Aslak, U.} \&
  \bibinfo{author}{Lehmann, S.}
\newblock \bibinfo{title}{The scales of human mobility}.
\newblock \emph{\bibinfo{journal}{Nature}} \textbf{\bibinfo{volume}{587}},
  \bibinfo{pages}{402--407} (\bibinfo{year}{2020}).

\bibitem{er2020universal}
\bibinfo{author}{Er-Jian, L.} \& \bibinfo{author}{Xiao-Yong, Y.}
\newblock \bibinfo{title}{A universal opportunity model for human mobility}.
\newblock \emph{\bibinfo{journal}{Scientific Reports (Nature Publisher Group)}}
  \textbf{\bibinfo{volume}{10}} (\bibinfo{year}{2020}).

\bibitem{gallotti2016stochastic}
\bibinfo{author}{Gallotti, R.}, \bibinfo{author}{Bazzani, A.},
  \bibinfo{author}{Rambaldi, S.} \& \bibinfo{author}{Barthelemy, M.}
\newblock \bibinfo{title}{A stochastic model of randomly accelerated walkers
  for human mobility}.
\newblock \emph{\bibinfo{journal}{Nature communications}}
  \textbf{\bibinfo{volume}{7}}, \bibinfo{pages}{1--7} (\bibinfo{year}{2016}).

\bibitem{gillner1998navigation}
\bibinfo{author}{Gillner, S.} \& \bibinfo{author}{Mallot, H.~A.}
\newblock \bibinfo{title}{Navigation and acquisition of spatial knowledge in a
  virtual maze}.
\newblock \emph{\bibinfo{journal}{Journal of cognitive neuroscience}}
  \textbf{\bibinfo{volume}{10}}, \bibinfo{pages}{445--463}
  (\bibinfo{year}{1998}).

\bibitem{foo2005humans}
\bibinfo{author}{Foo, P.}, \bibinfo{author}{Warren, W.~H.},
  \bibinfo{author}{Duchon, A.} \& \bibinfo{author}{Tarr, M.~J.}
\newblock \bibinfo{title}{Do humans integrate routes into a cognitive map?
  map-versus landmark-based navigation of novel shortcuts.}
\newblock \emph{\bibinfo{journal}{Journal of Experimental Psychology: Learning,
  Memory, and Cognition}} \textbf{\bibinfo{volume}{31}}, \bibinfo{pages}{195}
  (\bibinfo{year}{2005}).

\bibitem{norman2005perception}
\bibinfo{author}{Norman, J.~F.}, \bibinfo{author}{Crabtree, C.~E.},
  \bibinfo{author}{Clayton, A.~M.} \& \bibinfo{author}{Norman, H.~F.}
\newblock \bibinfo{title}{The perception of distances and spatial relationships
  in natural outdoor environments}.
\newblock \emph{\bibinfo{journal}{Perception}} \textbf{\bibinfo{volume}{34}},
  \bibinfo{pages}{1315--1324} (\bibinfo{year}{2005}).

\bibitem{sun2010perception}
\bibinfo{author}{Sun, Y.} \& \bibinfo{author}{Wang, H.}
\newblock \bibinfo{title}{Perception of space by multiple intrinsic frames of
  reference}.
\newblock \emph{\bibinfo{journal}{PloS one}} \textbf{\bibinfo{volume}{5}}
  (\bibinfo{year}{2010}).

\bibitem{weisberg2016some}
\bibinfo{author}{Weisberg, S.~M.} \& \bibinfo{author}{Newcombe, N.~S.}
\newblock \bibinfo{title}{How do (some) people make a cognitive map? routes,
  places, and working memory.}
\newblock \emph{\bibinfo{journal}{Journal of Experimental Psychology: Learning,
  Memory, and Cognition}} \textbf{\bibinfo{volume}{42}}, \bibinfo{pages}{768}
  (\bibinfo{year}{2016}).

\bibitem{vuong2019no}
\bibinfo{author}{Vuong, J.}, \bibinfo{author}{Fitzgibbon, A.~W.} \&
  \bibinfo{author}{Glennerster, A.}
\newblock \bibinfo{title}{No single, stable 3d representation can explain
  pointing biases in a spatial updating task}.
\newblock \emph{\bibinfo{journal}{Scientific reports}}
  \textbf{\bibinfo{volume}{9}}, \bibinfo{pages}{1--13} (\bibinfo{year}{2019}).

\bibitem{becu2020age}
\bibinfo{author}{B{\'e}cu, M.} \emph{et~al.}
\newblock \bibinfo{title}{Age-related preference for geometric spatial cues
  during real-world navigation}.
\newblock \emph{\bibinfo{journal}{Nature Human Behaviour}}
  \textbf{\bibinfo{volume}{4}}, \bibinfo{pages}{88--99} (\bibinfo{year}{2020}).

\bibitem{van2020large}
\bibinfo{author}{van~der Ham, I.~J.}, \bibinfo{author}{Claessen, M.~H.},
  \bibinfo{author}{Evers, A.~W.} \& \bibinfo{author}{van~der Kuil, M.~N.}
\newblock \bibinfo{title}{Large-scale assessment of human navigation ability
  across the lifespan}.
\newblock \emph{\bibinfo{journal}{Scientific Reports}}
  \textbf{\bibinfo{volume}{10}}, \bibinfo{pages}{1--12} (\bibinfo{year}{2020}).

\bibitem{marshall2018mathematical}
\bibinfo{author}{Marshall, J.~M.} \emph{et~al.}
\newblock \bibinfo{title}{Mathematical models of human mobility of relevance to
  malaria transmission in africa}.
\newblock \emph{\bibinfo{journal}{Scientific reports}}
  \textbf{\bibinfo{volume}{8}}, \bibinfo{pages}{1--12} (\bibinfo{year}{2018}).

\bibitem{yan2017universal}
\bibinfo{author}{Yan, X.-Y.}, \bibinfo{author}{Wang, W.-X.},
  \bibinfo{author}{Gao, Z.-Y.} \& \bibinfo{author}{Lai, Y.-C.}
\newblock \bibinfo{title}{Universal model of individual and population mobility
  on diverse spatial scales}.
\newblock \emph{\bibinfo{journal}{Nature communications}}
  \textbf{\bibinfo{volume}{8}}, \bibinfo{pages}{1--9} (\bibinfo{year}{2017}).

\bibitem{yan2019destination}
\bibinfo{author}{Yan, X.-Y.} \& \bibinfo{author}{Zhou, T.}
\newblock \bibinfo{title}{Destination choice game: A spatial interaction theory
  on human mobility}.
\newblock \emph{\bibinfo{journal}{Scientific reports}}
  \textbf{\bibinfo{volume}{9}}, \bibinfo{pages}{1--9} (\bibinfo{year}{2019}).

\bibitem{coutrot2019virtual}
\bibinfo{author}{Coutrot, A.} \emph{et~al.}
\newblock \bibinfo{title}{Virtual navigation tested on a mobile app is
  predictive of real-world wayfinding navigation performance}.
\newblock \emph{\bibinfo{journal}{PloS one}} \textbf{\bibinfo{volume}{14}},
  \bibinfo{pages}{e0213272} (\bibinfo{year}{2019}).

\bibitem{manley2015shortest}
\bibinfo{author}{Manley, E.}, \bibinfo{author}{Addison, J.} \&
  \bibinfo{author}{Cheng, T.}
\newblock \bibinfo{title}{Shortest path or anchor-based route choice: a
  large-scale empirical analysis of minicab routing in london}.
\newblock \emph{\bibinfo{journal}{Journal of transport geography}}
  \textbf{\bibinfo{volume}{43}}, \bibinfo{pages}{123--139}
  (\bibinfo{year}{2015}).

\bibitem{Malleson2018}
\bibinfo{author}{Malleson, N.} \emph{et~al.}
\newblock \bibinfo{title}{The characteristics of asymmetric pedestrian
  behavior: A preliminary study using passive smartphone location data}.
\newblock \emph{\bibinfo{journal}{Transactions in GIS}}
  \textbf{\bibinfo{volume}{22}}, \bibinfo{pages}{616} (\bibinfo{year}{2018}).

\bibitem{Dijkstra1959}
\bibinfo{author}{Dijkstra, E.}
\newblock \bibinfo{title}{A note on two problems in connexion with graphs}.
\newblock \emph{\bibinfo{journal}{Numerische Mathematik}}
  \textbf{\bibinfo{volume}{1}}, \bibinfo{pages}{269} (\bibinfo{year}{1959}).

\bibitem{Fechner1860}
\bibinfo{author}{Fechner, G.~T.}
\newblock \emph{\bibinfo{title}{Elements of Psychophysics}}
  (\bibinfo{publisher}{Howes, D H; Boring, E G (eds.)}, \bibinfo{year}{1860}).

\bibitem{newcombe1999misestimations}
\bibinfo{author}{Newcombe, N.}, \bibinfo{author}{Huttenlocher, J.},
  \bibinfo{author}{Sandberg, E.}, \bibinfo{author}{Lie, E.} \&
  \bibinfo{author}{Johnson, S.}
\newblock \bibinfo{title}{What do misestimations and asymmetries in spatial
  judgement indicate about spatial representation?}
\newblock \emph{\bibinfo{journal}{Journal of Experimental Psychology: Learning,
  Memory, and Cognition}} \textbf{\bibinfo{volume}{25}}, \bibinfo{pages}{986}
  (\bibinfo{year}{1999}).

\bibitem{bailenson1998road}
\bibinfo{author}{Bailenson, J.~N.}, \bibinfo{author}{Shum, M.~S.} \&
  \bibinfo{author}{Uttal, D.~H.}
\newblock \bibinfo{title}{Road climbing: Principles governing asymmetric route
  choices on maps}.
\newblock \emph{\bibinfo{journal}{Journal of Environmental Psychology}}
  \textbf{\bibinfo{volume}{18}}, \bibinfo{pages}{251--264}
  (\bibinfo{year}{1998}).

\bibitem{bailenson2000initial}
\bibinfo{author}{Bailenson, J.~N.}, \bibinfo{author}{Shum, M.~S.} \&
  \bibinfo{author}{Uttal, D.~H.}
\newblock \bibinfo{title}{The initial segment strategy: A heuristic for route
  selection}.
\newblock \emph{\bibinfo{journal}{Memory \& Cognition}}
  \textbf{\bibinfo{volume}{28}}, \bibinfo{pages}{306--318}
  (\bibinfo{year}{2000}).

\bibitem{christenfeld1995choices}
\bibinfo{author}{Christenfeld, N.}
\newblock \bibinfo{title}{Choices from identical options}.
\newblock \emph{\bibinfo{journal}{Psychological Science}}
  \textbf{\bibinfo{volume}{6}}, \bibinfo{pages}{50--55} (\bibinfo{year}{1995}).

\bibitem{howard2014hippocampus}
\bibinfo{author}{Howard, L.~R.} \emph{et~al.}
\newblock \bibinfo{title}{The hippocampus and entorhinal cortex encode the path
  and euclidean distances to goals during navigation}.
\newblock \emph{\bibinfo{journal}{Current Biology}}
  \textbf{\bibinfo{volume}{24}}, \bibinfo{pages}{1331--1340}
  (\bibinfo{year}{2014}).

\bibitem{marchette2014anchoring}
\bibinfo{author}{Marchette, S.~A.}, \bibinfo{author}{Vass, L.~K.},
  \bibinfo{author}{Ryan, J.} \& \bibinfo{author}{Epstein, R.~A.}
\newblock \bibinfo{title}{Anchoring the neural compass: coding of local spatial
  reference frames in human medial parietal lobe}.
\newblock \emph{\bibinfo{journal}{Nature neuroscience}}
  \textbf{\bibinfo{volume}{17}}, \bibinfo{pages}{1598} (\bibinfo{year}{2014}).

\bibitem{collett2004animal}
\bibinfo{author}{Collett, T.~S.} \& \bibinfo{author}{Graham, P.}
\newblock \bibinfo{title}{Animal navigation: path integration, visual landmarks
  and cognitive maps}.
\newblock \emph{\bibinfo{journal}{Current Biology}}
  \textbf{\bibinfo{volume}{14}}, \bibinfo{pages}{R475--R477}
  (\bibinfo{year}{2004}).

\bibitem{hafting2005microstructure}
\bibinfo{author}{Hafting, T.}, \bibinfo{author}{Fyhn, M.},
  \bibinfo{author}{Molden, S.}, \bibinfo{author}{Moser, M.-B.} \&
  \bibinfo{author}{Moser, E.~I.}
\newblock \bibinfo{title}{Microstructure of a spatial map in the entorhinal
  cortex}.
\newblock \emph{\bibinfo{journal}{Nature}} \textbf{\bibinfo{volume}{436}},
  \bibinfo{pages}{801--806} (\bibinfo{year}{2005}).

\bibitem{de2017spatial}
\bibinfo{author}{de~Cothi, W.} \& \bibinfo{author}{Spiers, H.~J.}
\newblock \bibinfo{title}{Spatial cognition: Goal-vector cells in the bat
  hippocampus}.
\newblock \emph{\bibinfo{journal}{Current Biology}}
  \textbf{\bibinfo{volume}{27}}, \bibinfo{pages}{R239--R241}
  (\bibinfo{year}{2017}).

\bibitem{toledo2020cognitive}
\bibinfo{author}{Toledo, S.} \emph{et~al.}
\newblock \bibinfo{title}{Cognitive map--based navigation in wild bats revealed
  by a new high-throughput tracking system}.
\newblock \emph{\bibinfo{journal}{Science}} \textbf{\bibinfo{volume}{369}},
  \bibinfo{pages}{188--193} (\bibinfo{year}{2020}).

\bibitem{poucet1983route}
\bibinfo{author}{Poucet, B.}, \bibinfo{author}{Thinus-Blanc, C.} \&
  \bibinfo{author}{Chapuis, N.}
\newblock \bibinfo{title}{Route planning in cats, in relation to the visibility
  of the goal}.
\newblock \emph{\bibinfo{journal}{Animal Behaviour}}
  \textbf{\bibinfo{volume}{31}}, \bibinfo{pages}{594--599}
  (\bibinfo{year}{1983}).

\bibitem{epstein2017cognitive}
\bibinfo{author}{Epstein, R.~A.}, \bibinfo{author}{Patai, E.~Z.},
  \bibinfo{author}{Julian, J.~B.} \& \bibinfo{author}{Spiers, H.~J.}
\newblock \bibinfo{title}{The cognitive map in humans: spatial navigation and
  beyond}.
\newblock \emph{\bibinfo{journal}{Nature neuroscience}}
  \textbf{\bibinfo{volume}{20}}, \bibinfo{pages}{1504} (\bibinfo{year}{2017}).

\bibitem{poulter2020vector}
\bibinfo{author}{Poulter, S.}, \bibinfo{author}{Lee, S.~A.},
  \bibinfo{author}{Dachtler, J.}, \bibinfo{author}{Wills, T.~J.} \&
  \bibinfo{author}{Lever, C.}
\newblock \bibinfo{title}{Vector trace cells in the subiculum of the
  hippocampal formation}.
\newblock \emph{\bibinfo{journal}{Nature Neuroscience}} \bibinfo{pages}{1--10}
  (\bibinfo{year}{2020}).

\bibitem{tolman1948cognitive}
\bibinfo{author}{Tolman, E.~C.}
\newblock \bibinfo{title}{Cognitive maps in rats and men.}
\newblock \emph{\bibinfo{journal}{Psychological review}}
  \textbf{\bibinfo{volume}{55}}, \bibinfo{pages}{189} (\bibinfo{year}{1948}).

\bibitem{fu2015single}
\bibinfo{author}{Fu, E.}, \bibinfo{author}{Bravo, M.} \&
  \bibinfo{author}{Roskos, B.}
\newblock \bibinfo{title}{Single-destination navigation in a
  multiple-destination environment: a new “later-destination attractor”
  bias in route choice}.
\newblock \emph{\bibinfo{journal}{Memory \& cognition}}
  \textbf{\bibinfo{volume}{43}}, \bibinfo{pages}{1043--1055}
  (\bibinfo{year}{2015}).

\bibitem{brunye2012planning}
\bibinfo{author}{Bruny{\'e}, T.~T.} \emph{et~al.}
\newblock \bibinfo{title}{Planning routes around the world: International
  evidence for southern route preferences}.
\newblock \emph{\bibinfo{journal}{Journal of Environmental Psychology}}
  \textbf{\bibinfo{volume}{32}}, \bibinfo{pages}{297--304}
  (\bibinfo{year}{2012}).

\bibitem{wilson1927probable}
\bibinfo{author}{Wilson, E.~B.}
\newblock \bibinfo{title}{Probable inference, the law of succession, and
  statistical inference}.
\newblock \emph{\bibinfo{journal}{Journal of the American Statistical
  Association}} \textbf{\bibinfo{volume}{22}}, \bibinfo{pages}{209--212}
  (\bibinfo{year}{1927}).

\bibitem{peer2020structuring}
\bibinfo{author}{Peer, M.}, \bibinfo{author}{Brunec, I.~K.},
  \bibinfo{author}{Newcombe, N.~S.} \& \bibinfo{author}{Epstein, R.~A.}
\newblock \bibinfo{title}{Structuring knowledge with cognitive maps and
  cognitive graphs}.
\newblock \emph{\bibinfo{journal}{Trends in Cognitive Sciences}}
  (\bibinfo{year}{2020}).

\bibitem{stern1988levels}
\bibinfo{author}{Stern, E.} \& \bibinfo{author}{Leiser, D.}
\newblock \bibinfo{title}{Levels of spatial knowledge and urban travel
  modeling}.
\newblock \emph{\bibinfo{journal}{Geographical Analysis}}
  \textbf{\bibinfo{volume}{20}}, \bibinfo{pages}{140--155}
  (\bibinfo{year}{1988}).

\bibitem{HMMmapmatching}
\bibinfo{author}{Newson, P.} \& \bibinfo{author}{Krumm, J.}
\newblock \bibinfo{title}{Hidden markov map matching through noise and
  sparseness}.
\newblock In \emph{\bibinfo{booktitle}{Proceedings of the 17th ACM SIGSPATIAL
  International Conference on Advances in Geographic Information Systems}}, GIS
  ’09, \bibinfo{pages}{336–343} (\bibinfo{publisher}{Association for
  Computing Machinery}, \bibinfo{address}{New York, NY, USA},
  \bibinfo{year}{2009}).
\newblock \urlprefix\url{https://doi.org/10.1145/1653771.1653818}.

\bibitem{DPAlgorithm}
\bibinfo{author}{Douglas, D.~H.} \& \bibinfo{author}{Peucker, T.~K.}
\newblock \bibinfo{title}{Algorithms for the reduction of the number of points
  required to represent a digitized line or its caricature}.
\newblock \emph{\bibinfo{journal}{Cartographica: the international journal for
  geographic information and geovisualization}} \textbf{\bibinfo{volume}{10}},
  \bibinfo{pages}{112--122} (\bibinfo{year}{1973}).

\bibitem{wilks1938large}
\bibinfo{author}{Wilks, S.~S.}
\newblock \bibinfo{title}{The large-sample distribution of the likelihood ratio
  for testing composite hypotheses}.
\newblock \emph{\bibinfo{journal}{The annals of mathematical statistics}}
  \textbf{\bibinfo{volume}{9}}, \bibinfo{pages}{60--62} (\bibinfo{year}{1938}).

\bibitem{zhang1993model}
\bibinfo{author}{Zhang, P.}
\newblock \bibinfo{title}{Model selection via multifold cross validation}.
\newblock \emph{\bibinfo{journal}{The annals of statistics}}
  \bibinfo{pages}{299--313} (\bibinfo{year}{1993}).

\bibitem{ZenodoDOI}
\bibinfo{author}{Buongiorno, C.} \emph{et~al.}
\newblock \bibinfo{title}{Pednav (1.1)}.
\newblock \emph{\bibinfo{journal}{Zenodo.
  https://doi.org/10.5281/zenodo.5187718}}  (\bibinfo{year}{2021}).

\end{thebibliography}

\begin{addendum}
\item 
P.S. and C.R. thank Amsterdam Institute for Advanced Metropolitan Solutions, Enel Foundation, DOVER, and all of the members of the MIT Senseable City Laboratory Consortium for supporting this research. This material is based upon work supported by the Center for Brains, Minds and Machines (CBMM), funded by NSF STC award CCF-1231216. C.B. and A.R. acknowledge the MISTI/MITOR fund. A. R. acknowledges Compagnia di San Paolo. 
This work was performed using HPC resources from the ``M\'esocentre'' computing center of CentraleSup\'elec and Ecole Normale Sup\'erieure Paris-Saclay supported by CNRS and R\'egion \^{I}le-de-France. Y.Z. acknowledges Prof. Bo Huang and the Chinese University of Hong Kong for supporting his academic visit at the MIT Senseable City Laboratory.

\item[Author Contributions]  A.R. and P.S. conceived and supervised the research. 
C.B. and Y.Z. led and performed the exploratory data analysis. 
C.B. designed the asymmetry testing procedure, the stochastic modeling, the validation method and performed the simulations. 
Y.Z. processed the data, conducted statistical analyses, and drafted the paper. M.K. and D.T. carried out exploratory modeling converging on the presented model;
M.K. drafted the introduction, content related to cognitive science and neuroscience, cognitive science research and methodology; 
C.R. contributed to conceptualise and design the research; 
J.T. provided conceptualization and feedback; all authors contributed to writing and revising the paper.

\item[Competing Interests] The authors declare that they have no
competing financial interests.
 \item[Correspondence] Correspondence and requests for materials
should be addressed to Paolo Santi~(email: psanti@mit.edu).
\end{addendum}

\end{document}


\maketitle

\section{Street Networks}

In Supplementary Figure \ref{fig:overview}, we present overview and zoomed-in vies of the street networks used in the study.

\begin{figure}[h!]
    \centering
    \includegraphics[width=\textwidth]{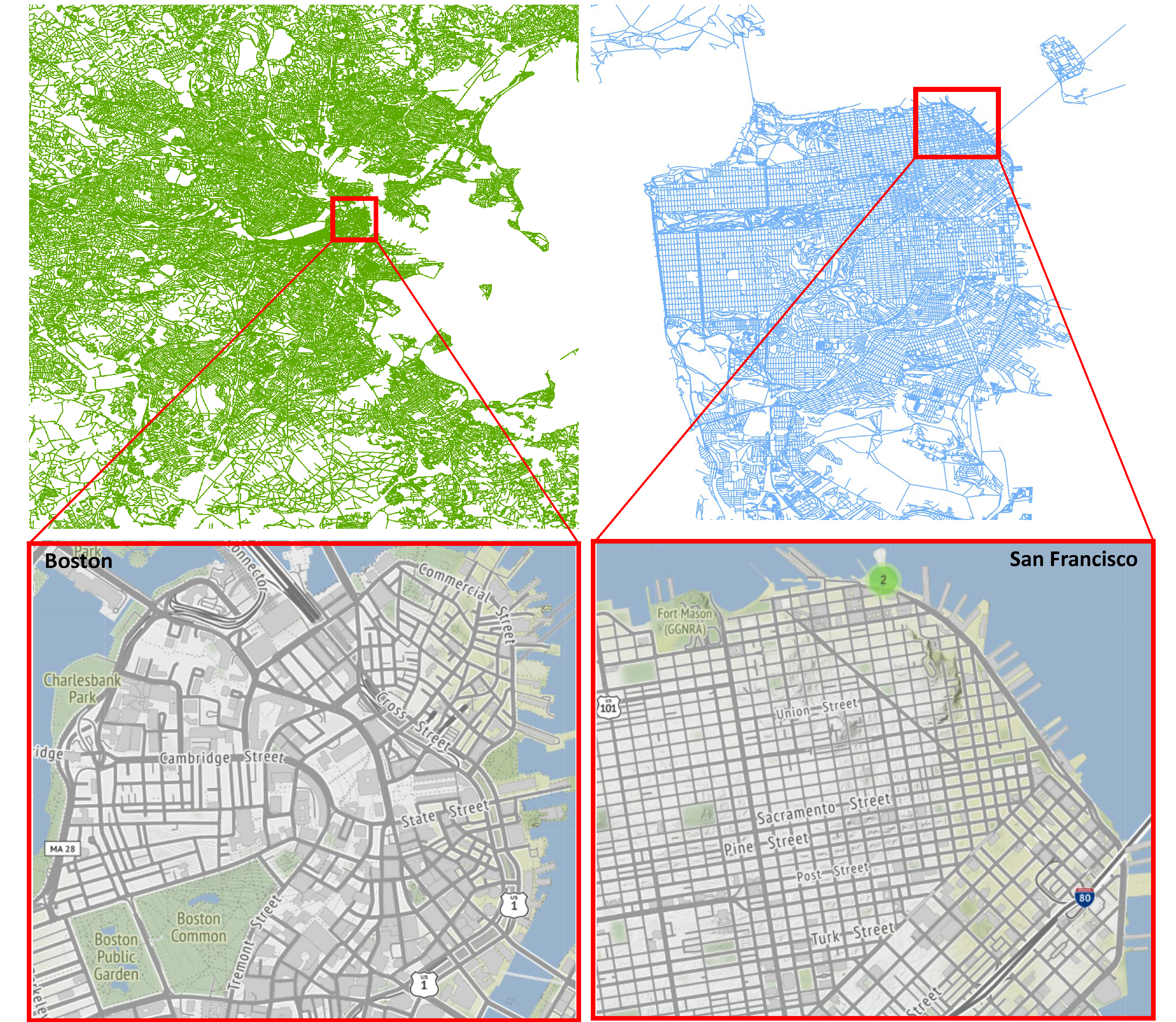}
    \caption{Overviews (upper) and zoomed-in views (lower) of the street network used in the analysis in Boston (right) and in San Francisco (left).}
    \label{fig:overview}
\end{figure}

\section{Walking Paths}
\subsection{Shortest Paths}
The summary statistics of the shortest paths in both Boston and San Francisco is shown in Supplementary Table \ref{tab:path-length}.

\begin{figure}[h!]
    \centering
    \begin{minipage}[b]{\textwidth}
        \includegraphics[width=\textwidth]{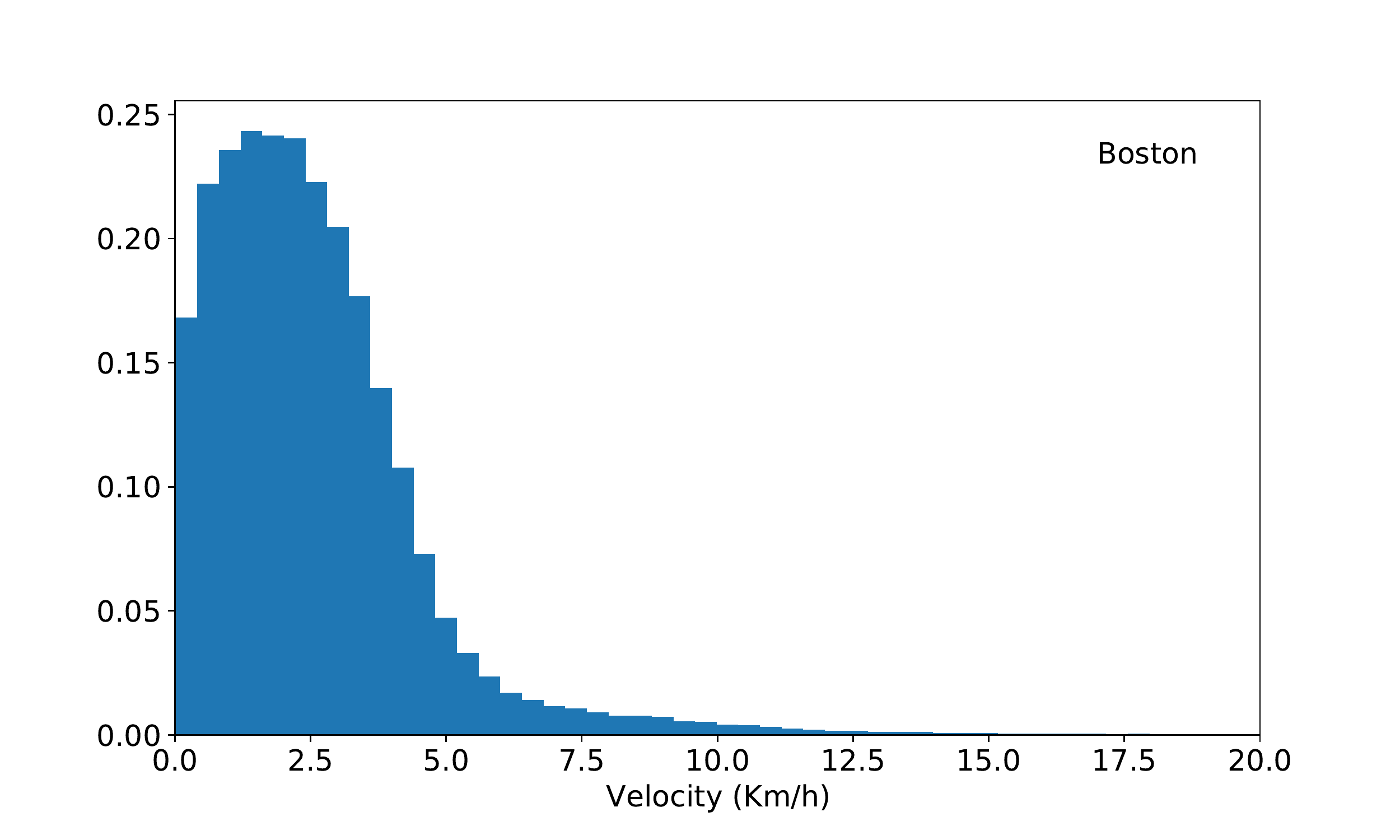}
    \end{minipage}
    \begin{minipage}[b]{\textwidth}
        \includegraphics[width=\textwidth]{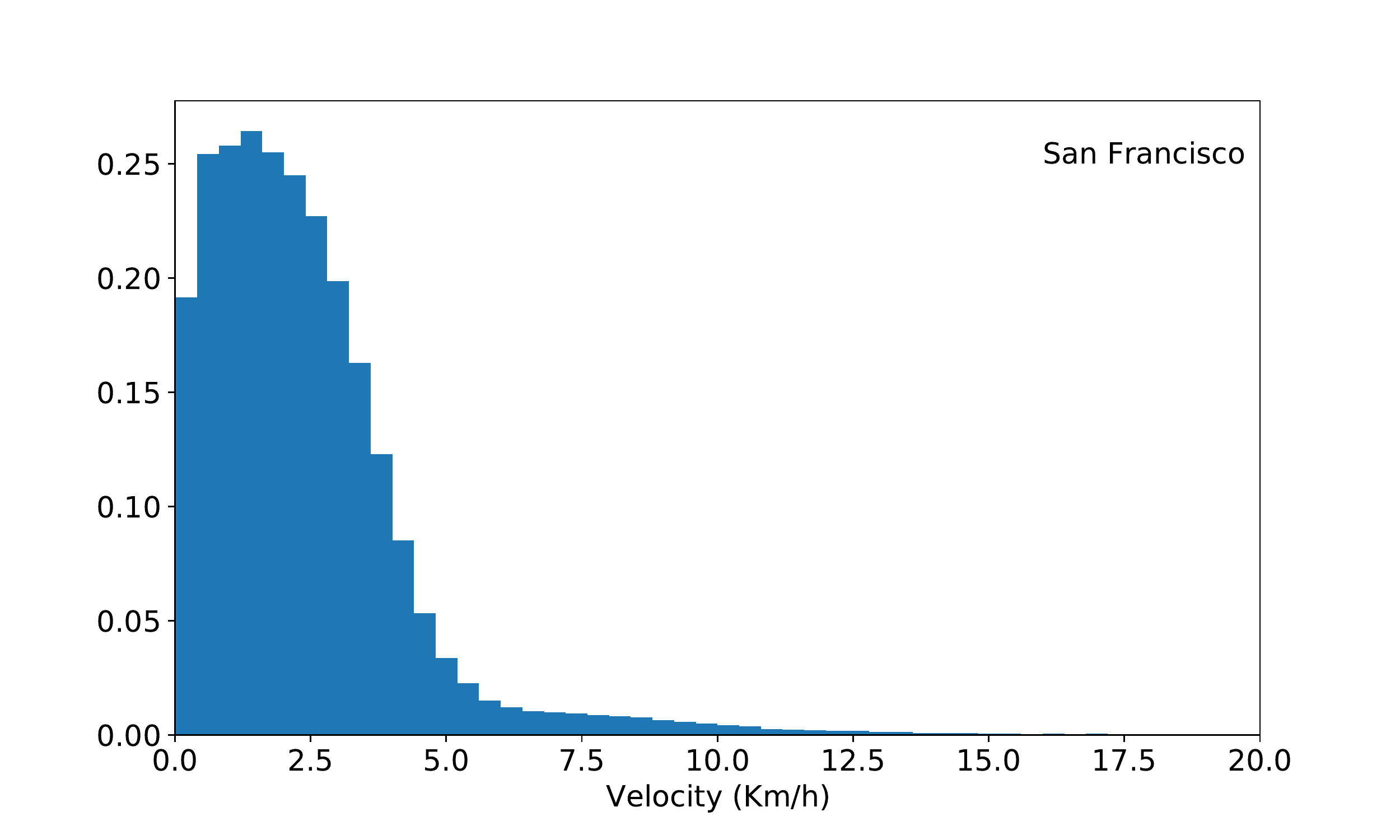}
    \end{minipage}
    
    \caption{Distribution of path velocity in Boston and San Francisco.}
    \label{fig:length_compare}
\end{figure}

\subsection{Human Paths}

The summary statistics of the human paths in both Boston and San Francisco is shown in Supplementary Table \ref{tab:path-length}.  Supplementary Figure 2 reports the trip velocity distribution extracted from data, in in Boston and San Francisco.

\begin{table}[h!]
\centering
\caption{ Summary statistics of human paths and shortest distance paths after data cleaning.}
\label{tab:path-length}
\resizebox{0.7\textwidth}{!}{%
\begin{tabular}{@{}|l|r|r|r|r|r|@{}}
\toprule
\multicolumn{2}{|l|}{\multirow{2}{*}{\textbf{Walking Paths}}} & \multicolumn{2}{c|}{\textbf{Boston}} & \multicolumn{2}{c|}{\textbf{San Francisco}} \\ \cmidrule(l){3-6} 
\multicolumn{2}{|l|}{}                           & Human   & Shortest & Human   & Shortest \\ \midrule
\multicolumn{2}{|l|}{\textbf{Count}}             & 165,645  & 165,645   & 189,075  & 189,075   \\ \midrule
\multirow{7}{*}{\textbf{Length (m)}} & \textbf{mean} & 856.0   & 758.0    & 868.1   & 781.7    \\ \cmidrule(l){2-6} 
                                 & \textbf{std}  & 843.6   & 718.3    & 912.1   & 796.0    \\ \cmidrule(l){2-6} 
                                 & \textbf{min}  & 200.0   & 200.0    & 200.0   & 200.0    \\ \cmidrule(l){2-6} 
                                 & \textbf{25\%}  & 372.9   & 345.9    & 363.7   & 341.9    \\ \cmidrule(l){2-6} 
                                 & \textbf{50\%}  & 596.4   & 536.9    & 583.7   & 535.3    \\ \cmidrule(l){2-6} 
                                 & \textbf{75\%}  & 1,019.7  & 899.0    & 1,017.4  & 914.9    \\ \cmidrule(l){2-6} 
                                 & \textbf{max}  & 23,167.1 & 18,396.8  & 36,377.7 & 29,176.8  \\ \bottomrule
\end{tabular}%
}
\end{table}

\begin{figure}[h!]
    \centering
    \begin{minipage}[b]{\textwidth}
        \includegraphics[width=\textwidth]{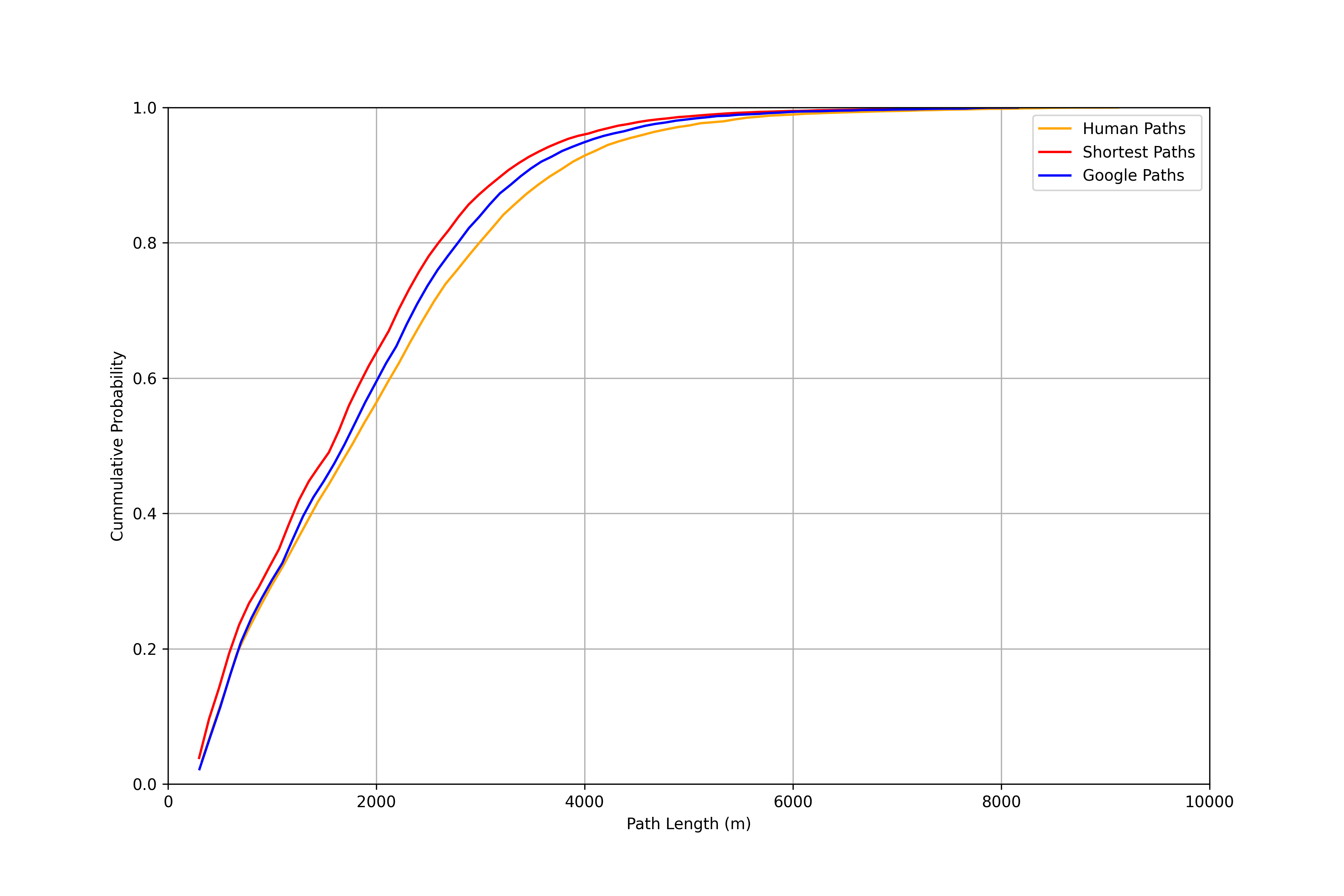}
    \end{minipage}
    \begin{minipage}[b]{\textwidth}
        \includegraphics[width=\textwidth]{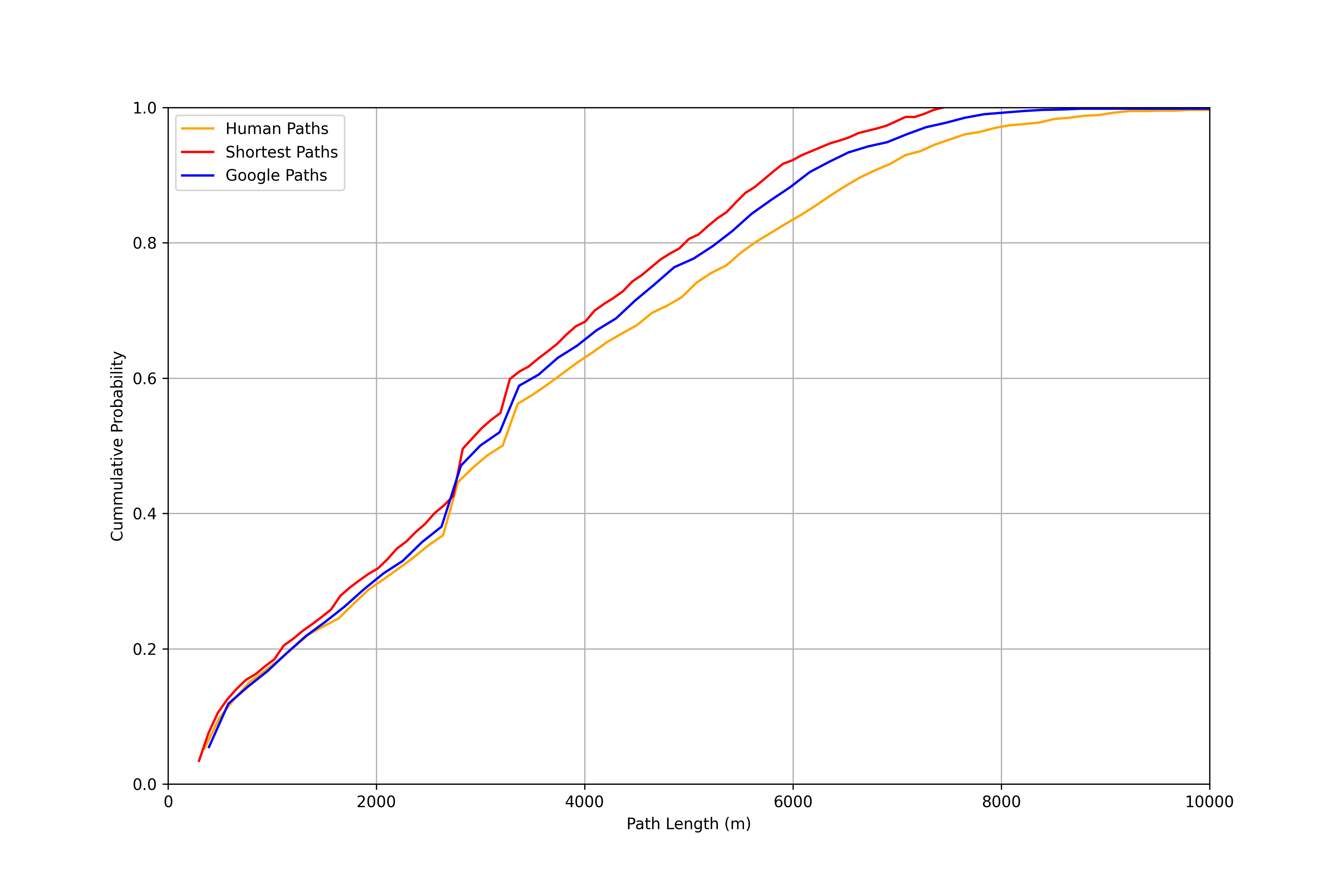}
    \end{minipage}
    
    \caption{Comparison among the cumulative densities of human, shortest, and Google path lengths in Boston (upper) and San Francisco (lower).}
    \label{fig:length_compare}
\end{figure}

\begin{table}[h!]
\centering
\caption{Length and geometric comparison results among human paths, shortest paths, and Google paths in Boston.}
\label{tab:geo-compare-bos}
\resizebox{\textwidth}{!}{%
\begin{tabular}{@{}|l|l|l|l|l|l|l|l|@{}}
\toprule
\multicolumn{1}{|c|}{\multirow{2}{*}{\textbf{Boston}}} &
  \multicolumn{3}{c|}{\textbf{Path Length (m)}} &
  \multicolumn{2}{c|}{\textbf{Jaccard Similarity}} &
  \multicolumn{2}{c|}{\textbf{Hausdorf Distance}} \\ \cmidrule(l){2-8} 
\multicolumn{1}{|c|}{} & Human  & Shortest & Google & H vs. G & H vs. S & H vs. G & H vs. S \\ \midrule
count                  & 10,853  & 10,853    & 9,254  & 10,853   & 10,853   & 10,853   & 10,853   \\ \midrule
mean                   & 1,950.9 & 1709.1   & 1,839.9 & 0.39    & 0.35    & 185.2   & 191.2   \\ \midrule
std                    & 1,324.2 & 1152.4   & 1,209.8 & 0.31    & 0.30    & 182.2   & 185.1   \\ \midrule
min                    & 214.1  & 202.6    & 202.6  & 0.00    & 0.00    & 0.0     & 0.0     \\ \midrule
25\%                   & 832.7  & 731.3    & 813.8  & 0.13    & 0.11    & 66.4    & 66.9    \\ \midrule
50\%                   & 1,764.7 & 1575.5   & 1,686.0 & 0.32    & 0.28    & 131.6   & 138.7   \\ \midrule
75\%                   & 2,724.3 & 2382.5   & 2,548.4 & 0.63    & 0.54    & 246.2   & 256.6   \\ \midrule
max                    & 9,119.9 & 7868.9   & 8,159.6 & 1.00    & 1.00    & 1,560.8  & 2,080.7  \\ \bottomrule
\end{tabular}%
}
\end{table}

\begin{table}[h!]
\centering
\caption{Length and geometric comparison results among human paths, shortest paths, and Google paths in San Francisco.}
\label{tab:geo-compare-sf}
\resizebox{\textwidth}{!}{%
\begin{tabular}{@{}|l|l|l|l|l|l|l|l|@{}}
\toprule
\multicolumn{1}{|c|}{\multirow{2}{*}{\textbf{San Francisco}}} &
  \multicolumn{3}{c|}{\textbf{Path Length (m)}} &
  \multicolumn{2}{c|}{\textbf{Jaccard   Similarity}} &
  \multicolumn{2}{c|}{\textbf{Hausdorff   Distance}} \\ \cmidrule(l){2-8} 
\multicolumn{1}{|c|}{} & Human   & Shortest & Google  & H vs. G & H vs. S & H vs. G & H vs. S \\ \midrule
count                  & 1,719    & 1,719     & 1,492    & 1,719    & 1,719    & 1,719    & 1,719    \\ \midrule
mean                   & 3,476.8  & 3,073.6   & 3,271.6  & 0.39    & 0.38    & 293.4   & 311.3   \\ \midrule
std                    & 2,273.9  & 1,906.6   & 2,067.2  & 0.34    & 0.37    & 375.1   & 374.1   \\ \midrule
min                    & 207.3   & 207.3    & 207.3   & 0.00    & 0.00    & 0.0     & 0.0     \\ \midrule
25\%                   & 1,683.2  & 1,508.1   & 1,611.6  & 0.09    & 0.06    & 42.1    & 35.9    \\ \midrule
50\%                   & 3,211.5  & 2,861.9   & 2,997.6  & 0.30    & 0.24    & 161.8   & 192.2   \\ \midrule
75\%                   & 5,147.9  & 4,534.2   & 4,747.5  & 0.63    & 0.67    & 418.8   & 432.1   \\ \midrule
max                    & 11,659.7 & 7,441.9   & 15,092.9 & 1.00    & 1.00    & 3,591.5  & 2,427.8  \\ \bottomrule
\end{tabular}%
}
\end{table}

\section{Likelihood Cross-Validation}
A certain fraction of paths will have zero sample probability in our simulation, even though, in a hypothetical analytical derivation, no path would have a zero probability. Such zero probabilities need to be replaced by a small non-zero threshold $c$, because taking a logarithm of a zero results in infinite likelihood. Therefore, we associate path probability $c$ to each path with a sample probability smaller than $c$. To ensure the robustness of the results, we obtained qualitatively similar results with different values of $.000001 < c < .001$. 
Notably, given the uncontrolled set up of our data-generation procedure, this thresholding would be intrinsically necessary even using analytical probability derivation, since it is not possible to eliminate outliers in human paths. For example, an outlier could be a pedestrian who is having a detour to meet a friend, or see a shop. Obviously, no model can predict such cases.

\begin{figure}[h!]
    \centering
    \includegraphics[width=3.5cm]{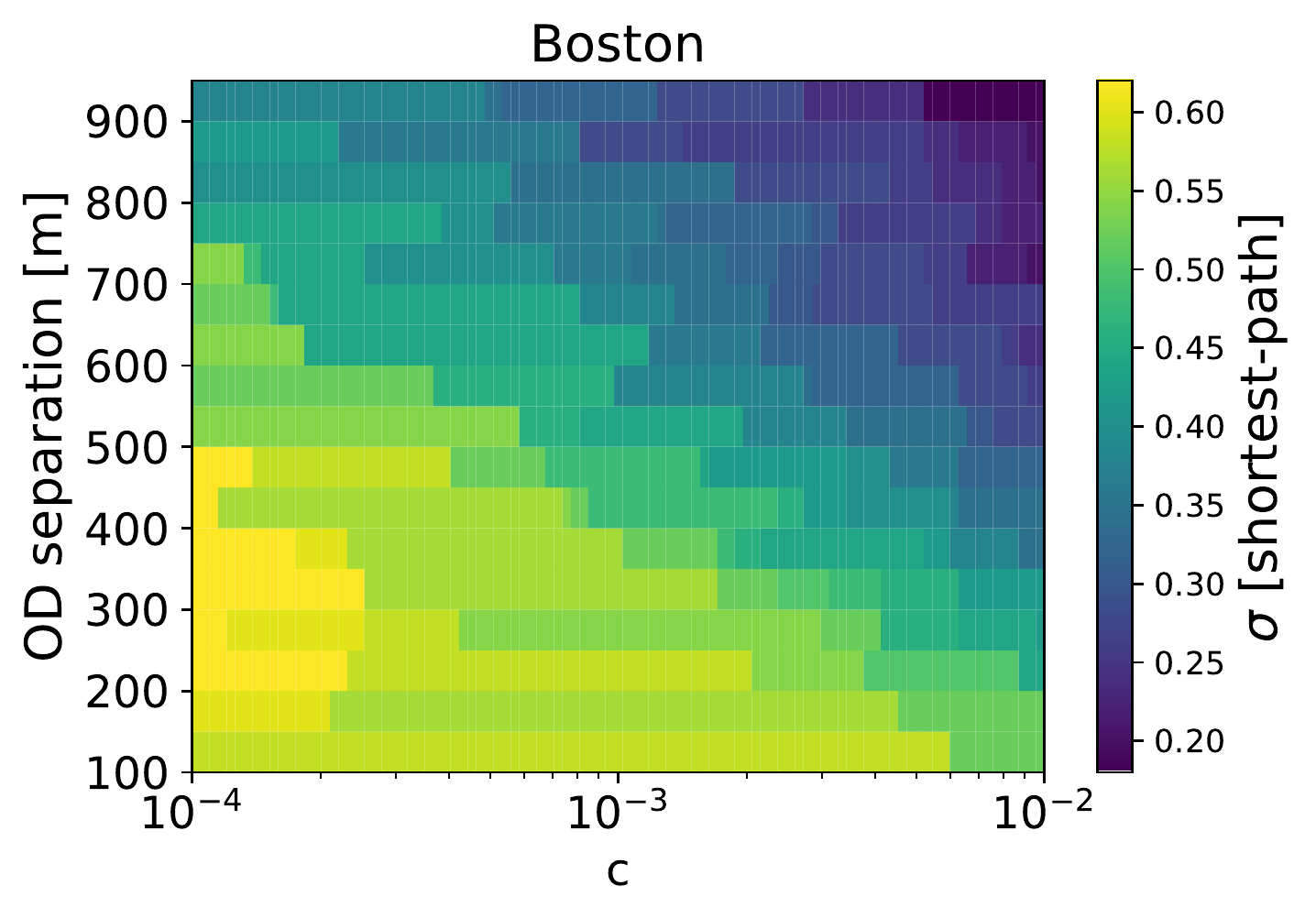}
    \includegraphics[width=3.5cm]{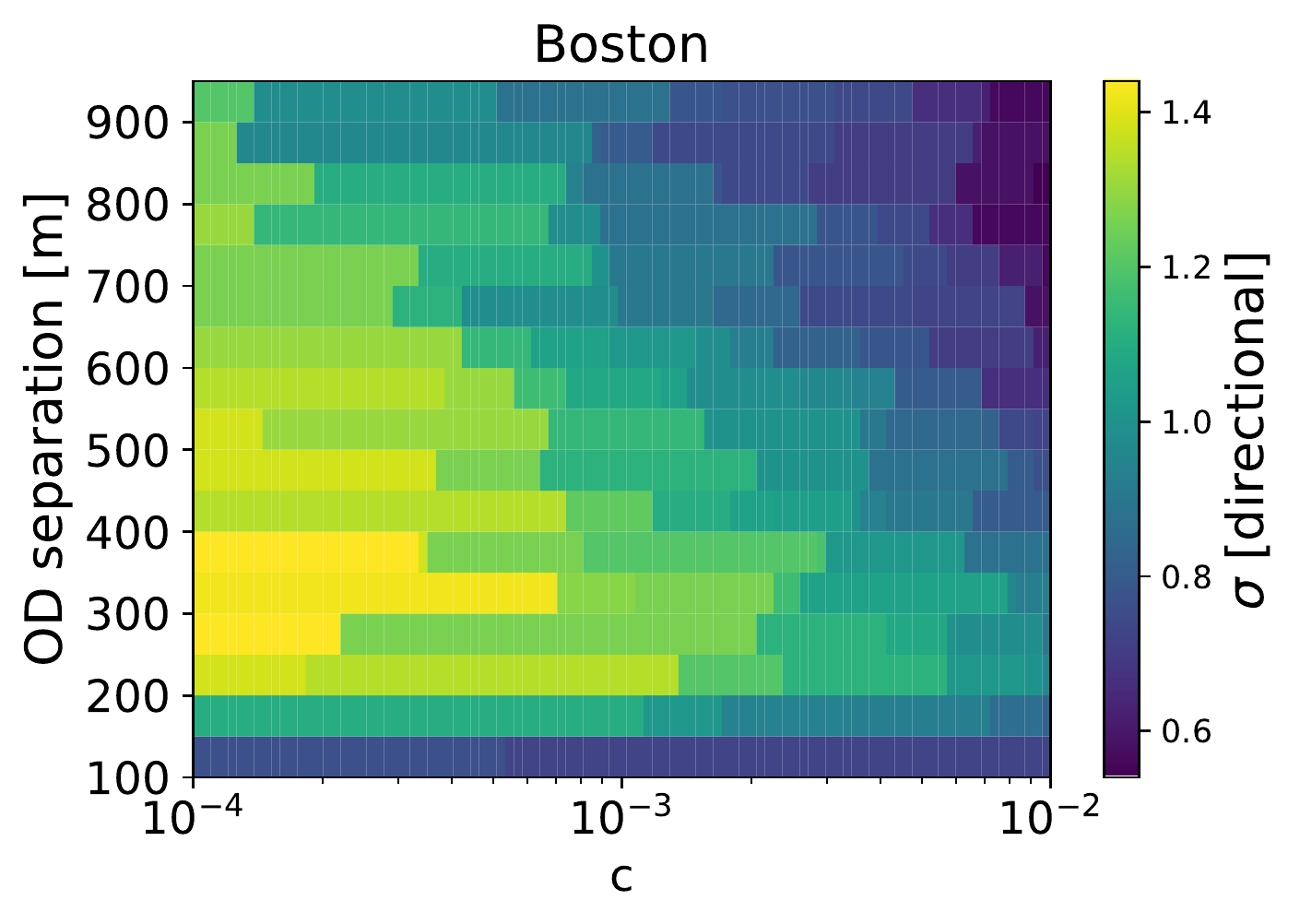}
    \includegraphics[width=3.5cm]{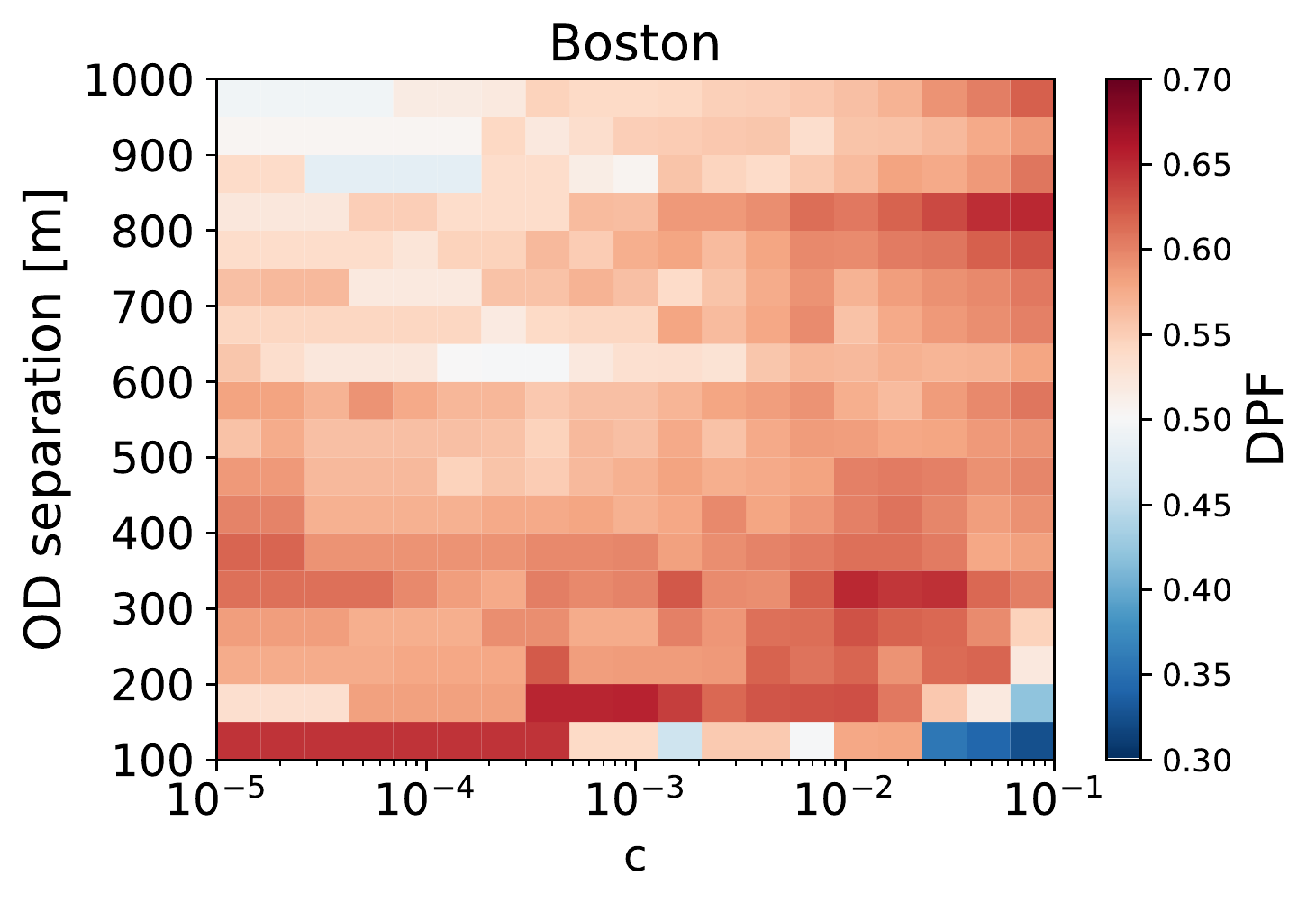}
    
    \includegraphics[width=3.5cm]{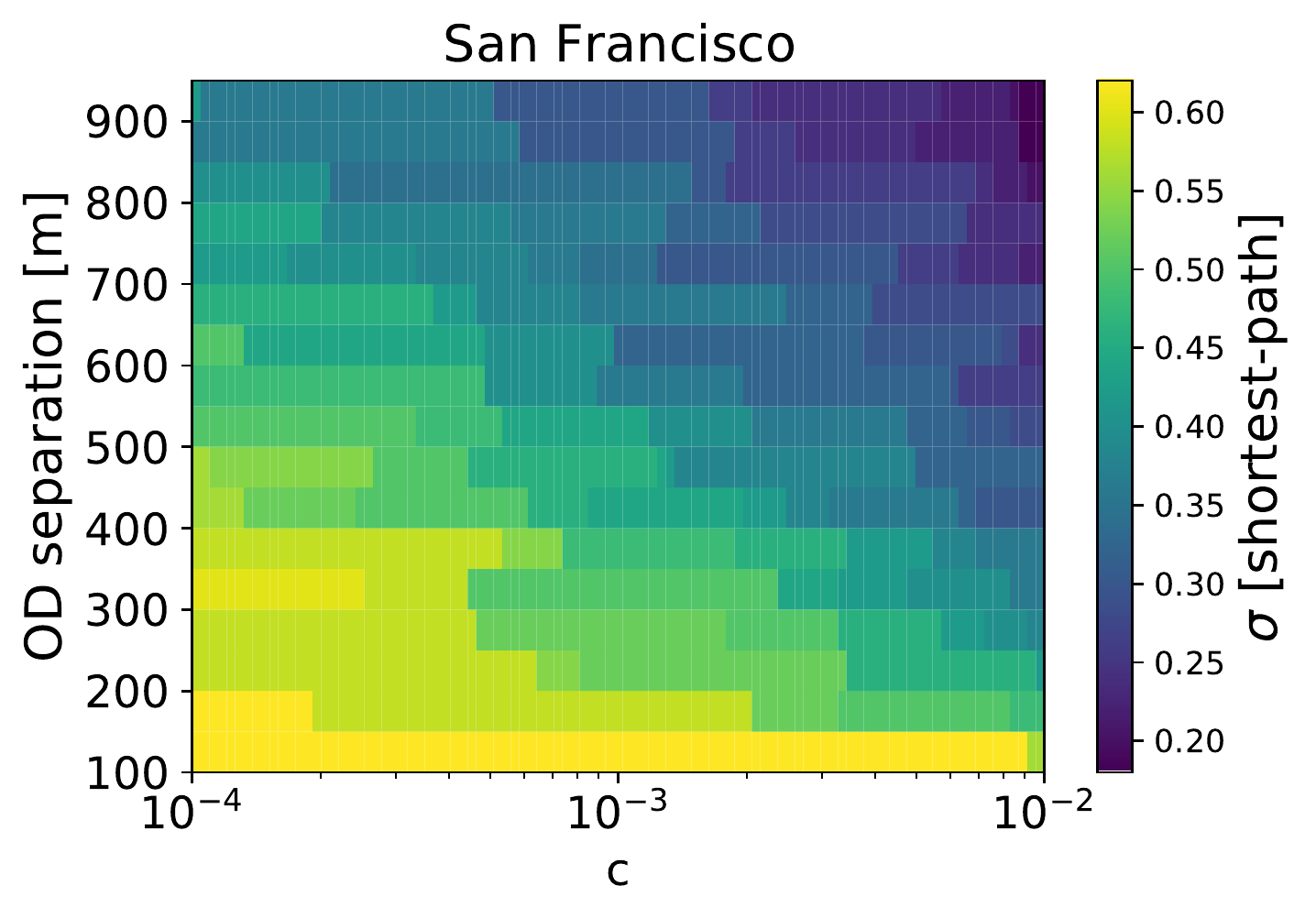}
    \includegraphics[width=3.5cm]{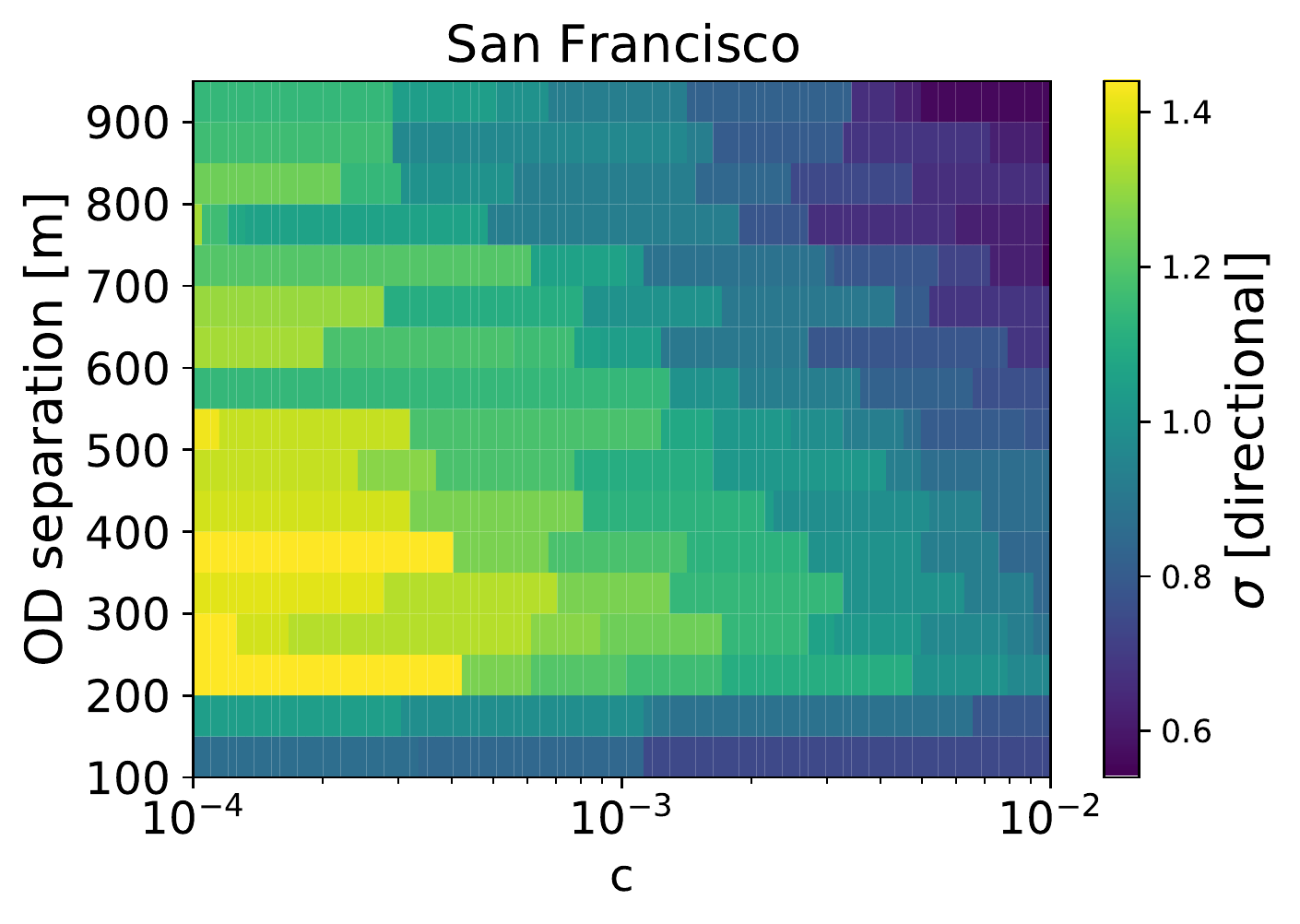}
    \includegraphics[width=3.5cm]{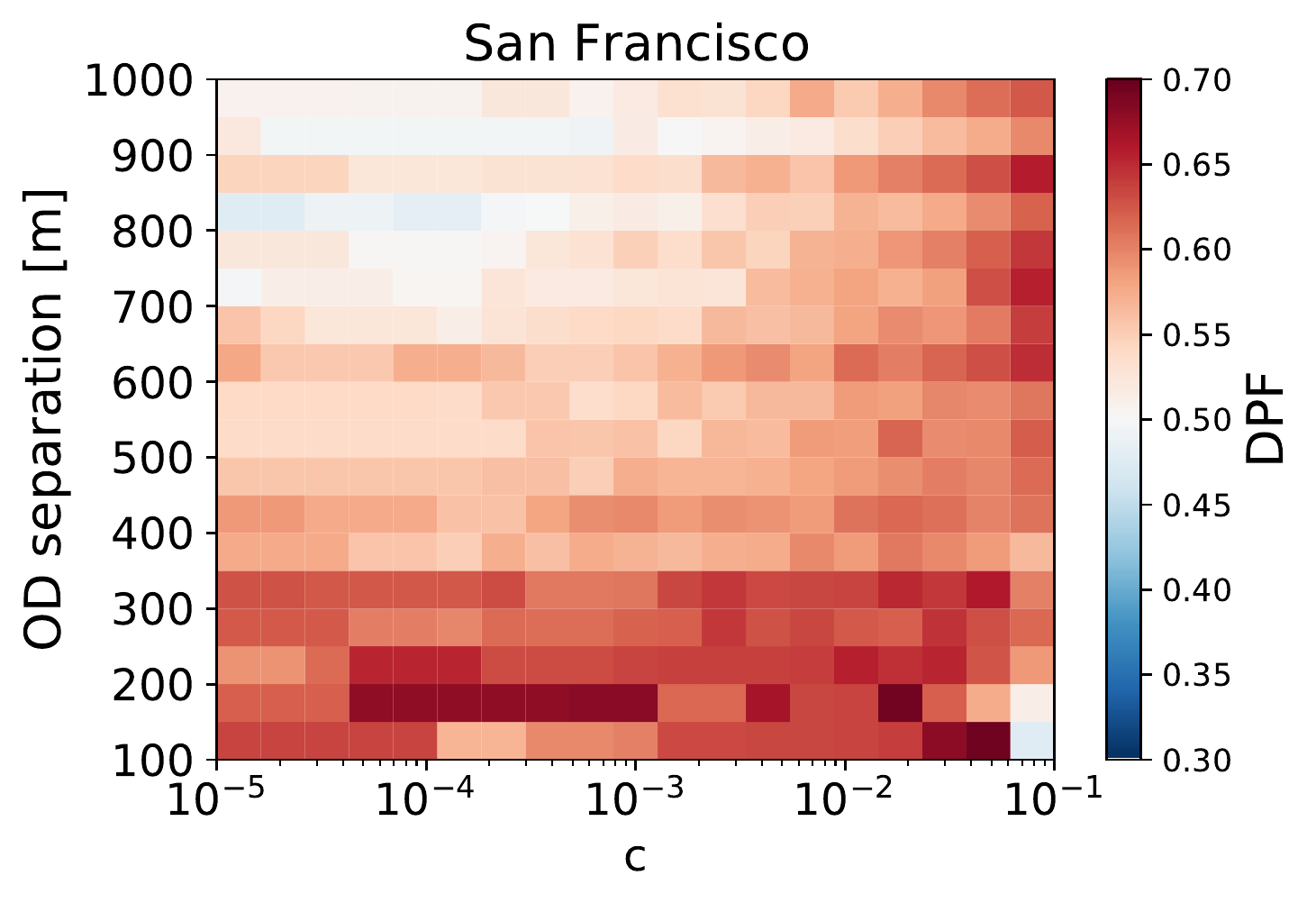}

    \caption{Probability threshold calibration. Upper panels refers to Boston, lower to San Francisco. Left panels shows the optimal $\sigma$ for the stochastic distance minimization model; central panels shows the optimal $\sigma$ for vector-based navigation, right panel shows the DPF in leave-on-out cross validation. All values are obtained for different $c$ and origin destination distance } 
    \label{fig:micro}
\end{figure}

\begin{figure}
    \centering
    \includegraphics[width=5.5cm]{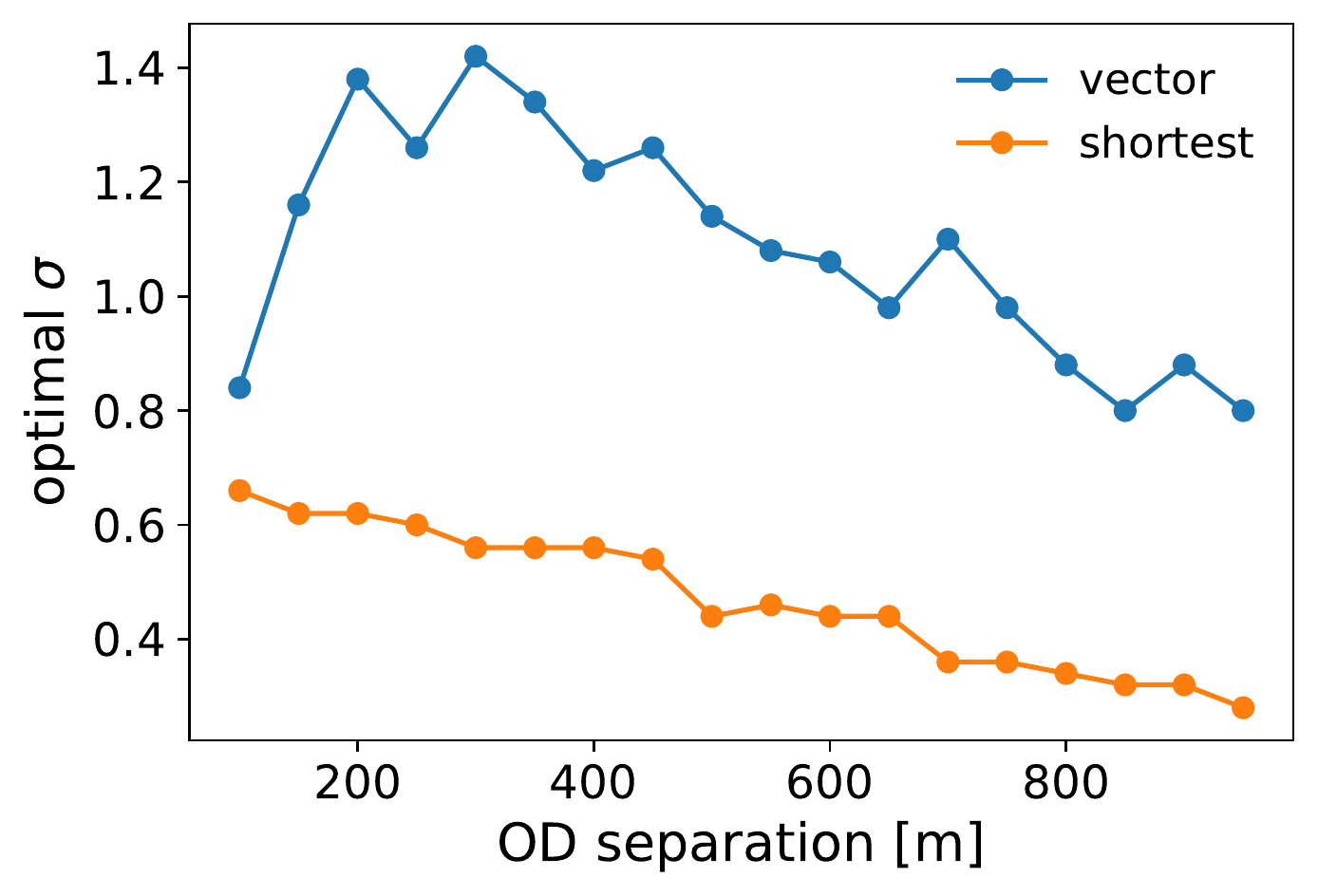}
    \includegraphics[width=5.5cm]{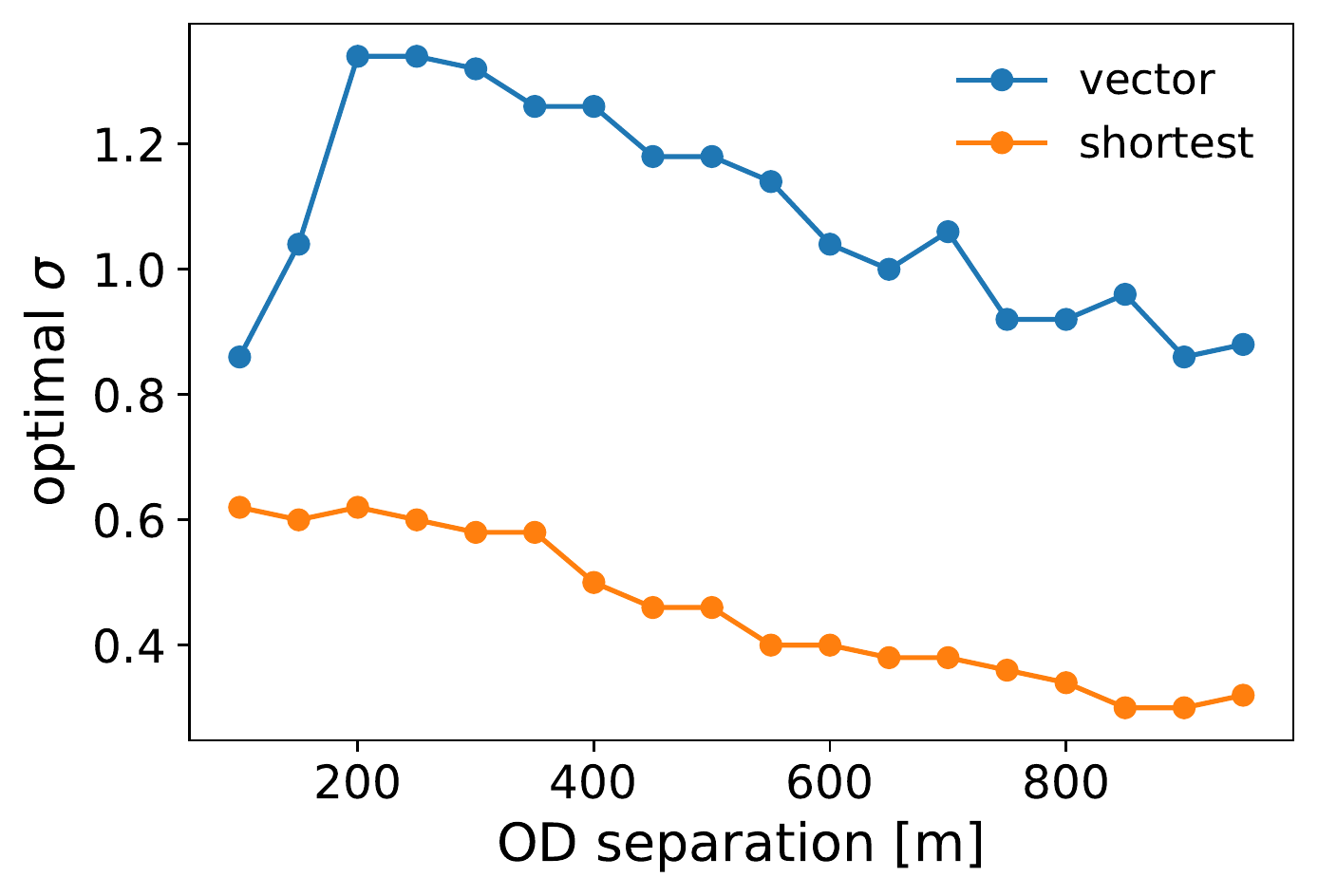}
    \caption{Optimal sigma for $c=.001$ as a function of the OD separation. Left plot refers to Boston; right plot refers to San Francisco.}
    \label{fig:optsimgaslice}
\end{figure}

The contour plots in Supplementary Figure~\ref{fig:micro} show the dependence of the parameter $\sigma$ on the OD separation and $c$. The results are qualitatively similar in both cities. When $c$ is low, the optimal $\sigma$ must be high. This is quite reasonable; in fact, to obtain non-zero probabilities for the outliers, the objective costs of the paths must be strongly perturbed.\\
Surprisingly, we can see that $\sigma$ decreases as the OD distance increases, which may be related to the traveling budget-time (Supplementary Figure~\ref{fig:optsimgaslice}). Whereas pedestrians could treat the shorter paths in a more leisurely fashion,  the longer paths are more likely premeditated, hence lower $\sigma$. 

The rightmost panels in Supplementary Figure~\ref{fig:micro} show the DPF in leave-one-out cross-validation obtained for different OD separation and $c$. This analysis shows that DPF is weakly affected by $c$, and that $DPF > 0.5$ for a wide set of values.

\section{Pseudo-code of Navigation Algorithms}

A pseudo-code description of the navigation algorithms is reported below.

\begin{algorithm}[h]
\caption{Calculate Stochastic shortest path}

\KwData{Network, origin, destination, $\sigma$}

\ForAll{Edge \in Network.Edges}{

  $Edge.Length$ := $Sample( exp ( \mathcal{N}( log(Edges.Length), \sigma^2 ) ))$ 
  
}

path : = Dijkstra(Network, origin, destination)

\algorithmicreturn{ path }

\end{algorithm}

\begin{algorithm}[h]
\caption{Calculate Vector-based path}

\KwData{Network, origin, destination, $\sigma$}

Let $dNetwork$ be a directed form of $Network$

\ForAll{Edge \in dNetwork.Edges}{

  $\alpha := \angle (\overrightarrow{Edge}, < \overrightarrow{Start(Edge), destination}> )$
  
  $Edge.Length := Sample( exp ( \mathcal{N}( log(\alpha ~Edges.Length), \sigma^2 )))$ 
  
}

path : = Dijkstra(dNetwork, origin, destination)

\algorithmicreturn{ path }
\end{algorithm}

\section{Additional Analysis}
\subsection{Individual Performances}
In this section we explore to what extent individuals display different performances in their ability of finding a shortest-path to destination. In particular, for each individual we aim to measure the fraction of times he/she correctly identifies the shortest path. Such fraction, computed for the specific individual, can then be compared to the average fraction computed across the entire population, and thus be considered as a metric of individual performance in finding shortest paths. 

In order to have sufficient statistical accuracy, we restricted our analysis to the $616$ individuals for which we have at least 50 paths recorded in the data set.

At the aggregate level of the entire population of 616 individuals, we have that $33\%$ of times the paths chosen equals the shortest path. However, at individual level we observe a large variation around this average value. In principle, the fraction of paths equal to the shortest path for each individual could be compared with the average value and we could perform a statistical test to check whether the observed deviation from a binomial distribution is significant. However, the observed deviation might be biased since not all the individuals have the same length distribution in their path sets.  As shown in Supplementary Figure~\ref{fig:individualPerf} (a), individuals who under-perform are characterized by a set of path with longer lengths than the control, and, of course, longer paths have lower probability to match exactly the shortest path. To account for this, we must control the bias introduced by differences in path length. This type of problem is typically addressed with matching set theory, that in a multi-dimensional setting would require the definition of a propensity score. However, being in our case the confounding variable only the length, the match can be addressed directly on this variable.  Specifically, to asses if a certain individual has a higher (or lower) fraction of path equal to the shortest, its sample fraction cannot be compared directly with average fraction of 33\% compute across the entire population; rather, we should select a tailored control set that match approximately the path length distribution of the tested individual. To do that, we binned the length of the path in steps of $50m$, and for each individual we sampled randomly a set of path that match the count of path on each bin of the individual under study. From such random sample, we obtain the control proportion of path equal to the shortest path. We repeated such a random sample 100,000 times. In Supplementary Figure~~\ref{fig:individualPerf} (c),(d),(e), we report the outcome of this process for three individuals who under-perform (c), perform the same (d), or over-perform (e) with respect to the null distribution. The null distribution converges to a Normal being the sum of independent Bernoulli trials, therefore in Supplementary Figure~\ref{fig:individualPerf} (f), we can use a z-score to highlight the presence of outliers. After this matching in Supplementary Figure~\ref{fig:individualPerf} (b), we show two users, one that under-performs and another that out-performs and both show to be unbiased with respect to the length distribution.


\begin{figure}
    \centering
    \includegraphics[width=\textwidth]{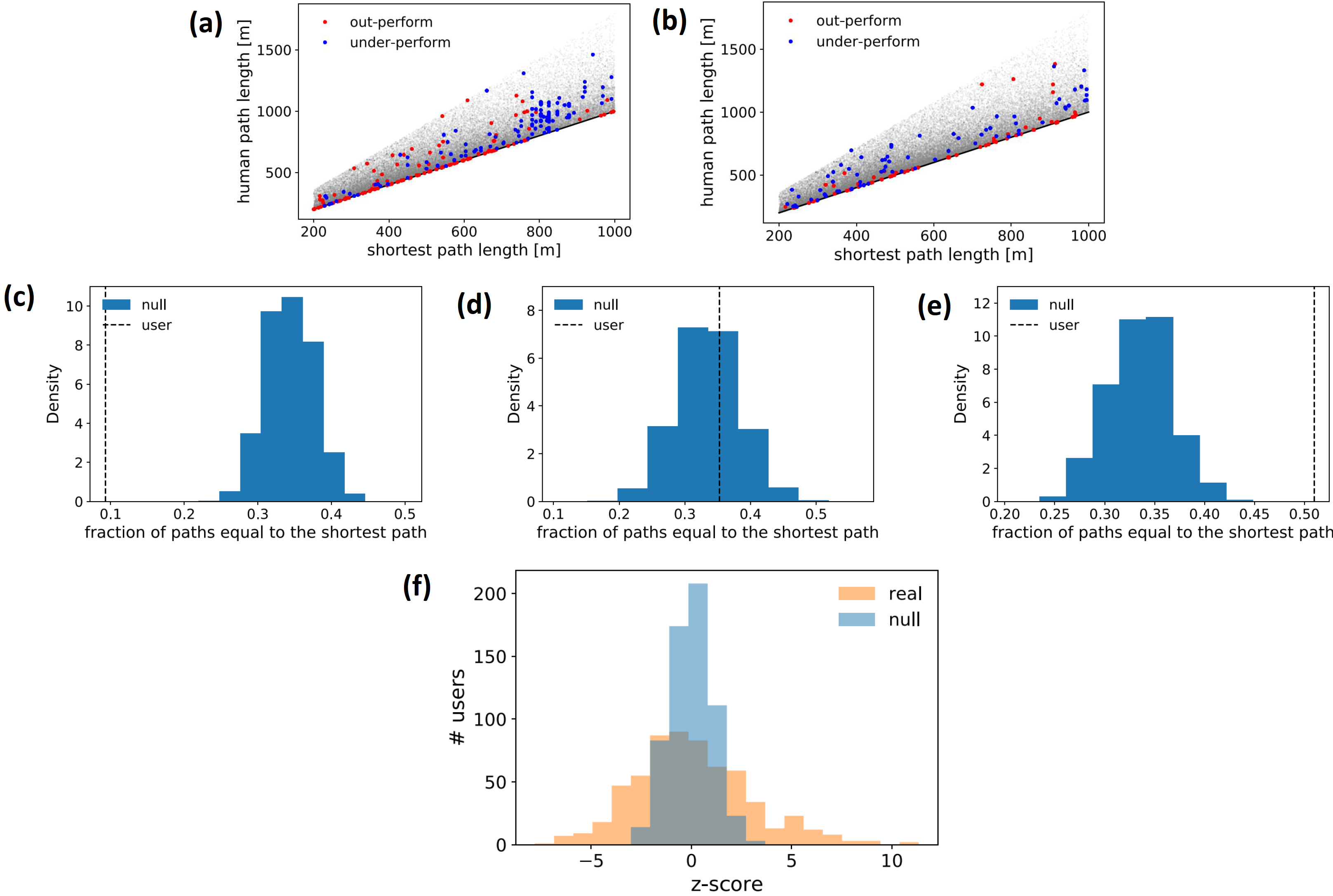}
    
    \caption{(a) each dots correspond to a path; red dots are the paths of an individual whose fraction of paths equal to the shortest path is significantly lower than the average fraction of 33\%; blue dots represent an individual whose fraction of path equal to the shortest path is significantly higher than 33\%. Figure (b) shows other two individuals who under-perform and out-perform after bias in path length distribution has been removed through matching. Figures (c),(d),(e) are three example of individuals with corresponding proportion of matching the shortest path (dotted line), compared with the expected proportion obtained from a control that match their length distribution. Figure(f) shows the z-score of each individual, showing that the distribution of z-scores of individuals in the data set is broader that the expected distribution of z-score (in blue) in case of no outliers.}
    \label{fig:individualPerf}
\end{figure}

\subsection{First Segment Strategy}
A possible cause of the observed asymmetry in human paths might be the tendency of selecting a relatively long straight segment at the beginning of the path, a tendency that has been observed in the literature and referred to as the Initial Straightest Segment (ISS) strategy. In this section, we perform a statistical test to assess this hypothesis.

\begin{figure}
    \centering
    \includegraphics[width=\textwidth]{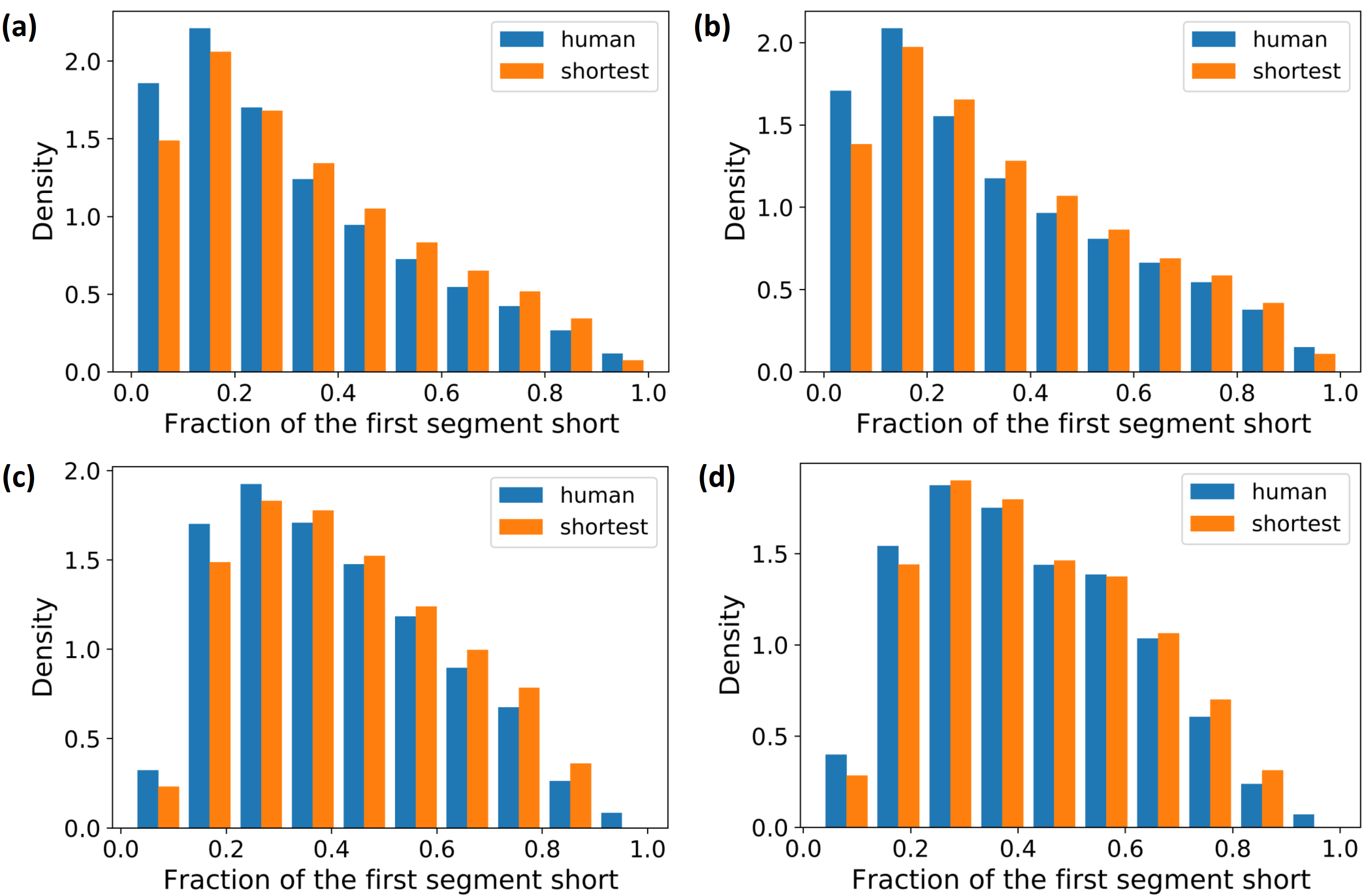}
       \caption{Distributions of the fraction of path length covered by the first segment in Boston $(a)$ $(c)$ and San Francisco $(b)$ $(d)$. The upper panels refers to all the paths, the lower panels are restricted to paths whose OD separation is at most 300m.}
    \label{fig:firstseg}
\end{figure}

In order to define the fist straight line sub-segment of each path, we simplified the trajectory with the DP algorithm with a cutoff of 30m. Then, we measured the path length for the first two points in of the DP-simplified representations. The average length covered by the first segment for humans in Boston is 219m while for the shortest path is 226m; for San Francisco both human and shortest path cover on average 256m. By looking at the fraction of the total path length covered by the first segment (Supplementary Figure~\ref{fig:firstseg}), in both cities we have observed that this fraction is larger for shortest than that of human paths.
To further confirm this observation, we also restricted our analysis to  paths with an OD separation of at most  300m. In this case, the absolute length covered by the first segment is slightly longer for the human paths (133m in both Boston and San Francisco) than for the shortest paths (130m). However, the fraction of the total path length covered by the first segment still shows higher value for the shortest path, as depicted in the lower panel of Supplementary Figure~\ref{fig:firstseg}.

To account for that, for the path with OD separation smaller than 300m we modeled straight-first segment propensity by setting a cost equal to zero to all straight segment that depart from each origin and searching for the shortest path. To do that, we need to identify the collections of street segments without a significant angle starting from the origin. This problem can be addressed by considering the dual edge-edge network. In this representation, a node is a street segment, and two street segments are linked if they are connected by a street intersection.  The link among two street intersections can be weighted by the angle among the two street. Therefore, in order to select the collection of all paths with a minimal or smooth angle from the origin, it is enough to cut all links (in the dual representation) with an angles higher than 20 degree and then find the set of street intersections that lie in a connected component that contains the origin node. However this approach fails in reproducing the slight increment in the first segment length observed on the human path: in fact, as result of this modeling the average total length covered by the first segment reaches 165m and 168m for Boston and San Francisco respectively, which are significantly longer than the observed ones.

\subsection{Decision Points}
In this section we explore the hypothesis that humans might have a tendency to minimize the number of decision points (road intersections) in their trajectory. To define such a number, we simplified each trajectory with DP algorithm with a 30m threshold. After that, the number of decision points is the total number of simplified segments minus one. We performed this analysis both on shortest paths and human paths for different OD separations. In upper panels of Supplementary Figure~\ref{fig:decision}, the average absolute number of decision points per trajectory is systematically higher on the human trajectory in both cities. We further computed the density of decision points since human paths are on average longer than shortest paths. The density of decision points were calculated as the number of decision points divided by the path length. In both cities, we reported a higher density of decision points for the humans than their shortest counterparts. We can then conclude that minimizing the number of decision points in the trajectory is likely not a significant factor in pedestrian path formation.

\begin{figure}
    \centering
    \includegraphics[width=\textwidth]{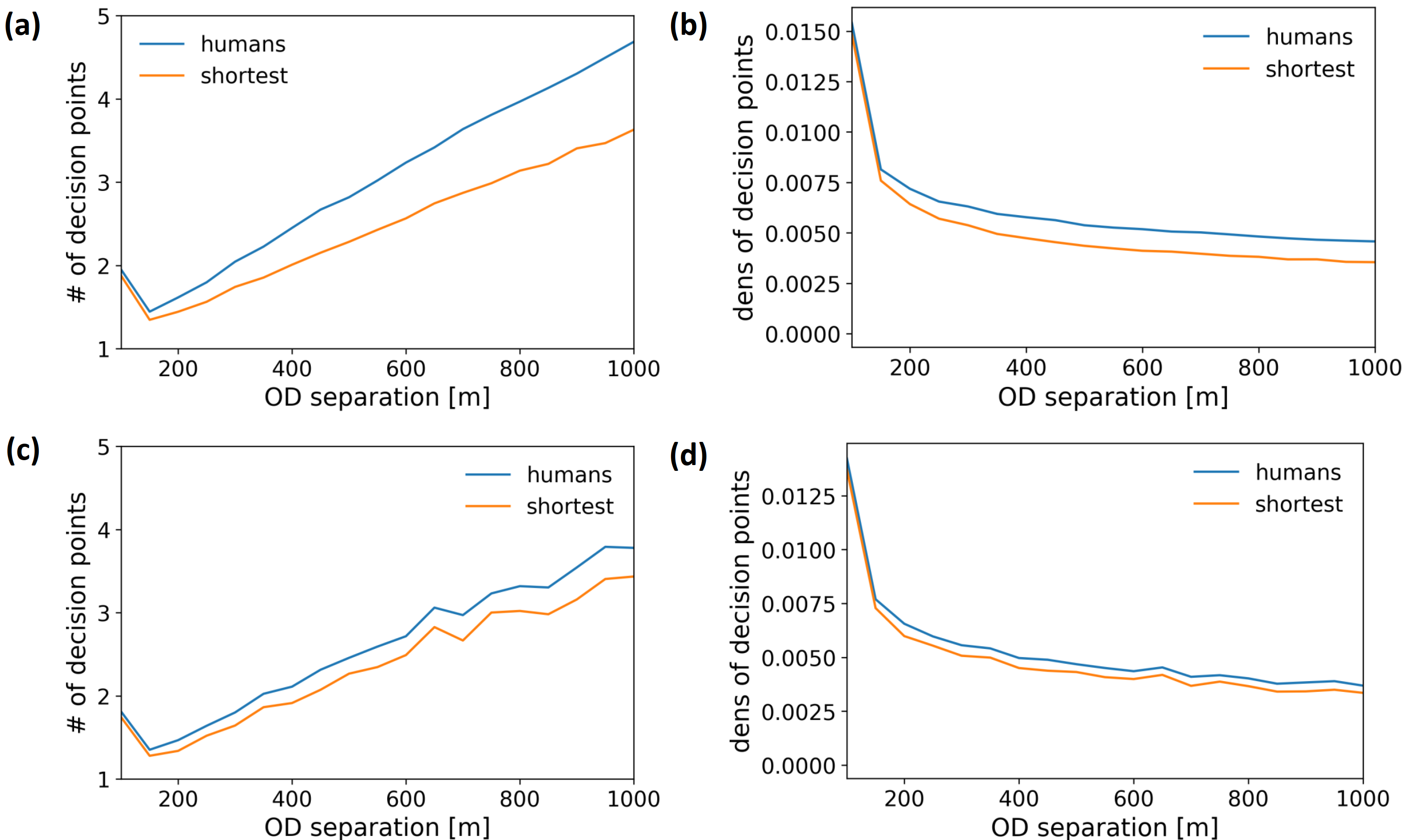}
       \caption{$(a)$ $(c)$ Average number of decision points as a function of the OD separation, $(b)$ $(d)$ density of decision points as a function of the the OD separation. $(a)$ and $(b)$ refer to Boston; $(c)$ and $(d)$ refer to San Francisco.}
    \label{fig:decision}
\end{figure}

\bibliographystyle{plain}